\documentclass[twocolumn,10pt]{article}

\usepackage{lipsum}
\usepackage{times}
\usepackage[margin=1in]{geometry}
\usepackage{amsmath, amssymb, amsfonts}
\usepackage{graphicx}
\usepackage{cite}
\usepackage{hyperref}
\usepackage{authblk}
\usepackage{setspace}
\usepackage{physics}
\usepackage{bm}
\usepackage{siunitx}
\usepackage{ragged2e}
\usepackage{float}
\usepackage{subcaption}
\usepackage{booktabs}
 \usepackage{multirow}

\begin{document}

\title{Disorder mediated fully compensated ferrimagnetic spin-gapless semiconducting behaviour in Cr$_3$Al Heusler alloy}%
\author{Reshna Elsa Philip$^{1}$, Pooja Vyas$^{2}$, Nikhil Joseph Joy$^{1}$, Sandip Kumar Kuila$^{3}$, Sonia Beniwal$^{4}$, Akshata Magar$^{5}$, Dinesh Kumar Shukla$^{6}$, Partha Pratim Jana$^{3}$, Amit Kumar$^{7,8}$, Aftab Alam$^{2}$, Jayakumar Balakrishnan$^{1}$, and Soham Manni$^{1\,*}$ \\
{\small $^{1}$Department of Physics, Indian Institute of Technology Palakkad, Kerala 678623, India \\ $^{2}$Department of Physics, Indian Institute of Technology Bombay, Mumbai 400076, India \\ $^{3}$Department of Chemistry, Indian Institute of Technology Kharagpur, West Bengal, India \\  $^{4}$Department of Physics, Central University of Haryana, Mahendergarh, India \\ $^{5}$School of Physics, Indian Institute of Science Education and Research, Thiruvananthapuram-695551, India \\ $^{6}$UGC-DAE Consortium for Scientific Research, Indore 452001, India \\ $^{7}$Solid State Physics Division, Bhabha Atomic Research Centre, Mumbai 400085, India \\ $^{8}$Homi Bhabha National Institute, Anushaktinagar, Mumbai 400094, India}}

\date{}

\twocolumn[
\maketitle

\begin{abstract}
Spin-gapless semiconductors (SGSs) that simultaneously host fully compensated ferrimagnetism are highly sought for energy-efficient and stray-field-free spintronic technologies, yet their realization in chemically disordered systems has remained elusive. Here, we demonstrate that the binary Heusler alloy Cr$_3$Al—despite adopting a fully A2-disordered structure—exhibits a rare coexistence of SGS transport and a fully compensated ferrimagnetic (FCF) ground state. Single-crystalline and polycrystalline Cr$_3$Al samples were synthesized, and comprehensive structural analyses using single-crystal XRD, synchrotron powder XRD, and neutron powder diffraction reveal complete Cr/Al site mixing. Remarkably, this chemical disorder does not disrupt magnetic order; instead, magnetization, X-ray magnetic circular dichroism (XMCD), and temperature-dependent neutron diffraction establish a robust compensated ferrimagnetic state with a vanishingly small ordered moment of $\sim$0.1(1)$\mu_B$/f.u and a high Curie temperature of 773 $\pm$ 2 K.
Electrical and thermal transport measurements uncover clear SGS characteristics, including weak temperature-dependent conductivity, very low Seebeck coefficients, and electron–hole–compensated transport. Hall measurements show unusual temperature-dependent carrier concentrations consistent with disorder-modified electronic states. First-principles calculations on an A2-disordered SQS structure reproduce the experimentally observed negligibly small magnetization (0.0072 $\mu_B$/f.u) and reveal a vanishing spin-up band gap—unambiguously supporting SGS behavior driven by chemical disorder.
Our results identify Cr$_3$Al as the first experimentally verified A2-disordered Heusler alloy exhibiting both fully compensated ferrimagnetism and spin-gapless semiconducting transport, positioning it as a robust and disorder-tolerant platform for next-generation, high-temperature spintronic devices.
\end{abstract}
\vspace{12pt}
]
\renewcommand{\thefootnote}{*}  
\footnotetext{Corresponding author\\\textit{Email address: smanni@iitpkd.ac.in}}
\renewcommand{\thefootnote}{\arabic{footnote}} 

\section{Introduction}
\begin{figure}[h!]
    \centering
    \includegraphics[width=0.99\linewidth]{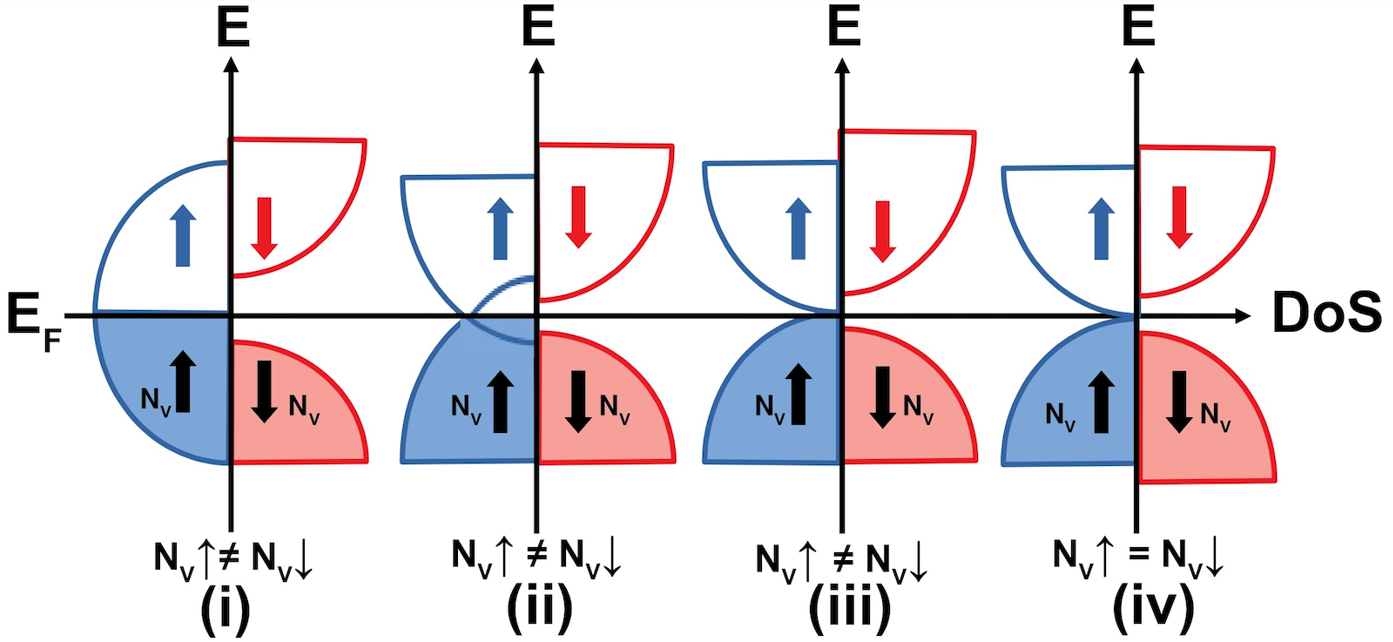}
    \caption{ Schematic density of states (DoS) for (i) HMF (ii) SSM (iii) SGS and (iv) fully compensated ferrimangetic (FCF) SGS, where $N_v\uparrow$ and $N_v\downarrow$ are the majority and minority spin valence electrons and $E_{\textit{F}}$ is the Fermi energy.  }
    \label{figure-1}
\end{figure}

Spintronics is a captivating field of research that exploits the electronic spin degrees of freedom for information storage and processing. Materials exhibiting high spin polarization are considered prime candidates for spintronic applications. Among these, Heusler alloys have shown great promise due to their highly tunable electronic and magnetic properties \cite{Tavares2023}. Groot \textit{et al.} predicted a distinct band structure in the half-Heusler compound NiMnSb, known as half metallic ferromagnetism (HMF) where the majority spin bands (usually referred as spin-up band) has a metallic behaviour while the minority spin band (referred as spin-down band) has a semiconducting gap at Fermi energy \cite{Groot1983}.  After the discovery of HMF \cite{Groot1983,Alijani2011,Seema2019}, different classes of materials with high spin polarization such as spin-gapless semiconductors (SGS) \cite{Wang2020,Xia2022,Gupta2024}, spin semimetals (SSM) \cite{Venkateswara_2023,Harikrishnan2023}, etc., are predicted from band structure calculations. For SSM, one of the spin channel is insulating/semiconducting and the other is semimetallic. The SGS are a class of materials in which, one spin channel exhibits a finite band gap while the other has a zero or nearly vanishing band gap.  As such, the excited charge carriers of both electrons and holes can be 100\% spin polarized with very high mobility. SGS materials are also highly sensitive to external factors such as pressure and magnetic fields \cite{Bainsla2015CoFecRGa,wang2008}. In addition to this, the spin injection efficiency in conventional ferromagnet or HMF/semiconductor interface, is highly limited by the Schmidt obstacle (due to conductivity mismatch) which can be effectively overcome by using SGS \cite{Xu_2015}. The primary experimental signatures of a SGS material are: (i) nearly temperature-independent conductivity, $\sigma$($T$), (ii) comparatively low and almost temperature-independent charge carrier concentration and (iii) low Seebeck coefficient ($|S|$) and coexistence of both electron and hole carriers \cite{Rani2020}. Further, SGS with a fully compensated ferrimagnetic (FCF) nature combine the distinct features of SGS with zero net magnetization, thereby effectively reducing energy losses in magnetic memory applications. Figure \ref{figure-1} illustrates the schematic density of states (DoS) for (i) HMF (ii) SSM (iii) SGS and (iv) fully compensated ferrimagnetic SGS.

 To realize such unique spin band structures, Heusler alloys have emerged as a versatile class of materials owing to their structurally simple and chemically tunable framework. The general chemical formula for full Heusler alloys is X\textsubscript{2}YZ, where Z is a main group element and X and Y are transition metals. These alloys typically crystallize in the $L$2\textsubscript{1} structure with space group $Fm\overline{3}m$. The unit cell of a full Heusler alloy consists of four interpenetrating face-centered cubic (FCC) sublattices, where the two X (X1 and X2), Y, and Z atoms occupy the Wyckoff positions  4$c$($\frac{1}{4}$,$\frac{1}{4}$,$\frac{1}{4}$), 4$d$($\frac{3}{4}$,$\frac{3}{4}$,$\frac{3}{4}$), 4$a$(0,0,0), and 4$b$($\frac{1}{2}$,$\frac{1}{2}$,$\frac{1}{2}$), respectively, in an ordered manner \cite{Wederni2024}. 4$c$ and 4$d$ sites are structurally equivalent and can be generally represented as 8$c$($\frac{1}{4}$,$\frac{1}{4}$,$\frac{1}{4}$).
In some systems, atoms at the octahedral sites (4$a$ and 4$b$) show a 50:50 disordered occupancy. This disorder arises from the similarity in the atomic size, electronegativity, and the structure factors of the constituent elements. Such partially disordered structures are referred to as B2-type. In cases where all four sites exhibit disordered occupancy, the structure is termed A2-type. When atoms X and Y in the X\textsubscript{2}YZ formula are the same transition metal, the structure simplifies to a binary DO$_3$-type alloy with the chemical formula X\textsubscript{3}Z \cite{Philip2024}. The crystal structures of the primitive ordered (DO$_3$) and A2-disordered  X\textsubscript{3}Z alloys are illustrated in Figure \ref{fi1}(a) and \ref{fi1}(b) of the supplementary information (SI), respectively.

 Compared to quaternary Heusler alloys, only a limited number of reports exist on binary DO$_3$-type alloys \cite{Wang2016,Zhang2012,Zhang2013,Jeong2012,Erkisi2018,Chattaraj2022,Gao2013,Li2009,Boekelheide2012Mar,Boekelheide2012Aug,Zhao2019,TOYOKI2021,Chakrabarti1971,Lind1984,DENBroeder1981}. Among these, Cr-based compounds are particularly underexplored \cite{Zhang2012,Chattaraj2022,Gao2013,Li2009,Boekelheide2012Mar,Boekelheide2012Aug,Zhao2019,TOYOKI2021,Chakrabarti1971,Lind1984,DENBroeder1981}. In 2013, Gao \textit{et al.} \cite{Gao2013} theoretically predicted that a Cr-based Heusler alloy, Cr$_3$Al can be tuned to SGS state from a metallic behavior by increasing the lattice constant from the equilibrium value of 5.92 Å to 6.22 Å within the DO\textsubscript{3} structure. They have also suggested that Cr$_3$Al exhibits ferrimagnetic ordering with a total magnetic moment of -3 $\mu$$_B$ per formula unit in the SGS state. Subsequently, a few studies on Cr$_3$Al thin films have reported commensurate antiferromagnetic ordering and semiconducting behavior under varying growth conditions \cite{Boekelheide2012Mar,Boekelheide2012Aug,Zhao2019,TOYOKI2021}. The Cr$_{1-x}$Al$_x$ alloys have been studied in bulk form for the past few decades. Compositions with approximately, $x = 0.25$ have been reported to exhibit pseudo-gap semiconducting behavior, characterized by a gap of around 60 meV with electrons as the majority charge carriers \cite{Chakrabarti1971}.  Lind \textit{et al.} \cite{Lind1984} found Cr$_3$Al alloy to possess commensurate antiferromagnetic ordering with semiconducting behaviour, using a detailed infrared reflectivity measurements. Selected area electron diffraction (SAED) analysis using transmission electron microscopy (TEM) revealed the presence of the X-phase ordering, characterized as either rhombohedral or disordered BCC, in Cr$_{100-x}$Al$_x$ alloys for $20\,\mathrm{at\%}\le x\le 25\,\mathrm{at\%}$ \cite{DENBroeder1981}. However, there are no thorough studies on the crystal structure, specially how it affects the magnetic ordering and electronic band structure of Cr$_3$Al alloy in bulk form.

In this communication, we report a comprehensive experimental and theoretical investigation of the electronic, magnetic, and transport properties of the bulk Cr$_3$Al polycrystal, including its first successful single-crystal synthesis. Structural analysis reveals that Cr$_3$Al does not crystallize in the ordered DO$_3$ structure commonly associated with Heusler-type compounds. Instead, it adopts a fully A2-disordered configuration, in which all atomic sites exhibit mixed Cr/Al occupancy. Remarkably, despite this pronounced disorder, Cr$_3$Al displays several intriguing and technologically relevant functionalities.
Magnetization measurements uncover a fully compensated ferrimagnetic (FCF) state, which is independently confirmed by first-principles simulations predicting an almost zero net magnetic moment. Temperature-dependent neutron diffraction further verifies the existence of long-range ferrimagnetic order with a compensated magnetic moment, underscoring the robustness of this magnetic ground state against chemical disorder.
Equally striking are the electronic and thermal transport responses. These measurements reveal characteristic signatures of spin-gapless semiconductor (SGS) behavior \cite{Venkateswara2018,Rani2019,Bainsla2015}. Corresponding electronic structure calculations for the A2-disordered phase reproduce these SGS-like features, establishing a consistent picture between experiment and theory. While SGS behavior has been extensively reported in ordered Heusler and related compounds, its realization in completely disordered alloys—particularly those exhibiting A2-type disorder—is exceedingly rare.
To the best of our knowledge, Cr$_3$Al represents the first experimentally verified A2-disordered alloy that simultaneously exhibits fully compensated ferrimagnetism and SGS-like transport characteristics. This unusual coexistence of disorder-tolerant magnetism and spin-selective transport positions Cr$_3$Al as a distinctive and promising platform for future spintronic technologies.

The rest of the paper is organized as follows. The experimental and theoretical details are discussed, in Sections 2 and 3 respectively.  Sections 4 and 5 are devoted to present the results and significant outcomes of this study, while Section 6 discusses the summary and main conclusions. 

\section{Experimental Section}
Polycrystalline Cr$_3$Al Heusler alloy sample is prepared with stoichiometric Chromium (Alfa Aesar, 99.99\%) and Aluminum (Alfa Aesar, 99.995\%) using an arc melting furnace under argon atmosphere. The melting process was repeated five times to obtain good compositional homogeneity. The final weight loss was less than 1\%. Cr$_3$Al single crystals were synthesized using metal flux method with Sn as the external flux. Cr, Al pieces and Sn granules were placed in an alumina crucible at a ratio of 3: 1: 80 and sealed in an evacuated quartz tube. The ampule was heated to 1100$^{\circ}$ C and held for 24 h for a homogeneous melt. The ampule was slowly cooled to 800$^{\circ}$ C at 3$^{\circ}$ C/hour. The single crystals were separated from the flux by centrifuging at this temperature. Submillimeter-sized single crystals were obtained using this growth technique (Figure \ref{structure}(b)).

Synchrotron-based  XRD measurements are carried out on polycrystalline powder samples using XRD beamline (BL-11) at the Indus-2 synchrotron source, Raja Ramanna Centre for Advanced Technology, Indore, India. Single-crystal X-ray diffraction (SCXRD) measurements were done on the Cr$_3$Al single crystals. The selected crystal was carefully mounted onto the top of goniometer head. The crystal was adhered to a nylon loop using paratone oil. The SCXRD data were collected at room temperature using a Bruker Photon II area detector with graphite-monochromator Mo-K$\alpha$ radiation ($\lambda$= 0.71073 \AA) with the help of Bruker D8 Quest four-circle diffractometer. The data collection, integration, and reduction were performed using the APEX 5 software package \cite{APEX5}.

 High temperature magnetization measurements were done using MPMS SQUID-VSM. X-ray magnetic circular dichroism (XMCD) experiments were performed at the XMCD beamline (BL-20) of Indus-2 Synchrotrons radiation source, RRCAT, Indore, in total electron yield mode. Circular polarization of X-rays were provided by the helical undulator installed at Indus-2 storage ring. The neutron powder diffraction patterns were recorded on the Powder Diffractometer-2 at BARC, Mumbai using neutrons of wavelength $\lambda$= $1.2443~\text{\AA}$ in the broad angular range of 3$^\circ$-137$^\circ$ using 5 linear position sensitive ${}^3\text{He}$ detectors. The temperature and field dependent electrical measurements were carried out using a Quantum Design Physical Property Measurement System (PPMS) and Oxford Teslatron PT-cryogen free superconducting magnet system. Measurements were performed in a magnetic field range of -7 T  to 7 T and in a temperature range of 2-300 K by sourcing a current of 1 mA. Thermopower measurements were performed utilizing a home-developed setup in a closed-cycle helium refrigerator (CCR) by employing controlled heat pulses and implementing a linear-fit method to reduce spurious voltages \cite{shukla2019}.

\section{Computational Section}
First-principles calculations were performed using spin-polarized density functional theory (DFT), as implemented in the Vienna ab initio Simulation Package (VASP) \cite{34_Kresse1996,35_Kresse1996,36_Kresse1993}, employing the projector augmented-wave (PAW) method \cite{37_Kresse1999}. The exchange-correlation (XC) functional was treated using a meta-generalized gradient approximation (meta-GGA), specifically the strongly constrained and appropriately normed (SCAN) functional \cite{38_Sun2015}. A plane-wave energy cutoff of 500 eV was used for wavefunction expansion.
Both ordered and disordered structures were fully relaxed until the total energy and atomic forces converged to 10\textsuperscript{-6} eV and 0.01 eV/\r{A}, respectively. For Brillouin zone (BZ) integration of the ordered Cr\textsubscript{3}Al structure, a 16\textsuperscript{3} Monkhorst-Pack k-point mesh \cite{39_Monkhorst1976} was used in conjunction with the tetrahedron method \cite{40_Blochl1994}.

\begin{figure*}[ht]
\hspace*{-0.48cm}\includegraphics[width=1.025\linewidth]{structural_details_1711_last.pdf}
\caption{\justifying(a) Reciprocal lattice reconstructions of the diffraction data of Cr$_3$Al single crystal. (b) As synthesized Cr$_3$Al single crystal. The Rietveld refinement of the room-temperature synchrotron XRD pattern of polycrystalline Cr$_3$Al sample, showing the fit with (c) FCC ($Fm\overline{3}m$) structure having A2-type disorder and (d) BCC ($Im\overline{3}m$) structure. The respective insets show the unit cells.}
\label{structure}
\end{figure*}
To model the A2-disordered phase, a 16-atom special quasirandom structure (SQS) \cite{41_Zunger1990} was generated using the mcsqs module of the ATAT code \cite{42_Vandewalle2013}. The SQS is an atomic configuration designed to statistically reproduce the correlation functions of a fully random alloy within a finite supercell, thereby capturing the essential features of chemical disorder.  For the disordered structure, BZ sampling was performed using a 12\textsuperscript{3} Monkhorst-Pack k-mesh within the tetrahedron method. Further, to compare the electronic band structure of the ordered and disordered phase, the band structure unfolding was performed using the 'easyunfold' package \cite{Zhu2024}

\section{EXPERIMENTAL RESULTS}
\subsection{Structural and Chemical characterization}
The comprehensive structural analysis of the single crystal and the polycrystalline alloy of Cr$_3$Al was carried out by single crystal XRD and synchrotron powder XRD, respectively. For the single crystal (shown in Figure \ref{structure}(b)), the reciprocal lattice reconstruction of the ($hk$0), ($h$0$l$), (0$kl$) planes from the single crystal XRD  data is presented in Figure \ref{structure}(a). The diffraction spots were indexed on the basis of  a body centred cubic lattice, since the sum of Miller indices $h$+$k$+$l$=2$n$. The structure was solved using the charge-flipping algorithm implemented in the program SUPERFLIP \cite{superflip}, and subsequently refined the data using JANA 2006 \cite{jana2006}. The EDS analysis reveals the atomic ratio between Cr and Al in the single crystal is approximately 3:1, the Cr position was mixed between Cr and Al (73$\pm$3\% Cr and 27$\pm$3\% Al). The structure determination using the charge-flipping algorithm converged reliably in the cubic space group $Im\overline{3}m$ (SG No:229) with $a$= 2.9422$\pm$0.0008 \AA. The solution generated only one Cr/Al mixed position 2$a$ (0,0,0). The details of refinement and crystallographic data are given in supplementary information (SI). A complete Cr/Al site disorder with nearly Cr:Al= 3:1 stoichiometry is obtained for the single crystal.

Compositional analysis on polycrystal using SEM-EDS revealed atomic percentages of approximately 77$\pm$2\% Cr and 23$\pm$7\% Al, confirming an atomic ratio close to 3:1. Crystal structure of the polycrystalline alloy is investigated using synchrotron powder XRD with $\lambda$=0.4746 \AA. The Rietveld refinement  of XRD data was done using Fullprof Suite software \cite{Reitveld} starting with the composition obtained from EDS measurement. The diffraction pattern is refined with the FCC structure ($Fm\overline{3}m$, SG No. 225), as shown in Figure \ref{structure}(c), and the data clearly indicate the absence of (111) and (200) superlattice reflections. This points to a similar chemical disorder in the polycrystalline sample also. To understand it further, we examined the structure factor of some of the important peaks. The Wyckoff positions of the atoms in DO$_3$-type X$_3$Z structure are: X at 8c (1/4,1/4,1/4), 4a (0,0,0), and Z at 4b (1/2,1/2,1/2). The structure factor for the X$_2$YZ in FCC structure is given by

{\scriptsize
\begin{equation}
F_{hkl} = 4\left( f_Z + f_Y e^{\pi i(h + k + l)} + f_X e^{\frac{\pi i}{2}(h + k + l)} + f_X e^{-\frac{\pi i}{2}(h + k + l)} \right)
\end{equation}
}

where $f_X$, $f_Y$, $f_Z$ are the atomic scattering factors of X, Y and Z atoms.  The intensities of each reflections are determined by the squares of the structure factors, given by the following expressions for different lattice reflections, 
\begin{equation}
F_{111} = 4(f_Z - f_Y) \\
\label{111eqn}
\end{equation}
\begin{equation}
F_{200} = 4(f_Z + f_Y - 2f_X) \\
\label{200eqn}
\end{equation}
\begin{equation}
F_{220} = 4(f_Z + f_Y + 2f_X) \\
\end{equation}
\begin{figure*}[ht]
\hspace*{-0.5cm}\includegraphics[width=1.04\linewidth]{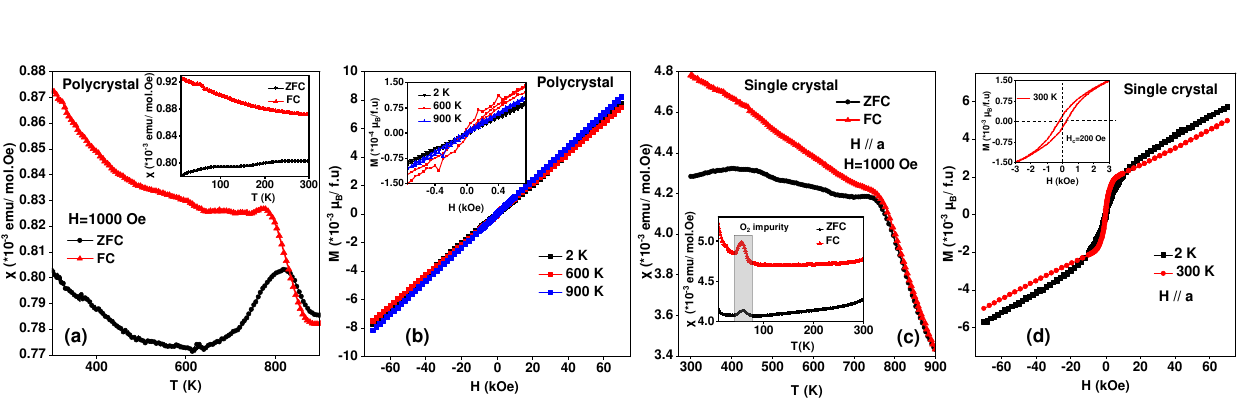}
\caption{\justifying Magnetic susceptibility ($\chi$) vs. temperature ($T$) of (a) polycrystal, (c) single crystal Cr$_3$Al from 300 to 900 K at 1000 Oe. The insets show the $T$-dependent magnetic susceptibility from 2 K to 300 K. (b) Magnetization ($M$) vs. magnetic field ($H$) of polycrystalline Cr$_3$Al at 2 K, 600 K and 900 K. The inset shows a zoomed-in view of the same. (d) $M$ vs. $H$ of single crystal at 2 K and 300 K. The inset shows a zoomed-in view of the same at 300 K, indicating the coercive field. }
	\label{magnetization}
\end{figure*}
$F_{111}$ and $F_{200}$ consist of the difference of atomic scattering factors and hence correspond to the order dependent reflections of the crystal lattice. Since $F_{220}$ contains sum terms, it is attributed to the order independent principal reflection. Therefore for an A2-disorderd structure where all atomic species are randomly distributed over all the available Wyckoff sites in the lattice, the (111) and (200) superlattice reflections vanishes completely \cite{Webster1969} as given by Eqs. \eqref{111eqn} and \eqref{200eqn}.
\begin{table}[H]
\small  
\renewcommand{\arraystretch}{1.09}  
\setlength{\tabcolsep}{0.6pt}  
\caption{\justifying\label{tab:table3}Structural data for single-crystal and polycrystalline Cr$_3$Al Heusler alloy.}
\centering
\begin{tabular}{l p{1.1cm} p{1.8cm} c c}
\hline 
Sample (Cr$_3$Al)\ \ \ & Crystal system &\  Space group & a$_{exp}$(\AA)  & \ \ \ \ $\chi^2$ \\                                                       
\hline
Single crystal &\ BCC &\ \ $Im\overline{3}m$ & 2.9422$\pm$0.0008 &\ \ \ \ 1.4\\  \\

Polycrystal & FCC-A2 disorder &\ \ $Fm\overline{3}m$ & 5.89$\pm$0.03 &\ \ \ \ 1.26\\

Polycrystal & BCC &\ \ $Im\overline{3}m$ & 2.95$\pm$0.05 &\ \ \ \ 2.4\\
\hline 
\end{tabular}
\end{table}
 The refinement in $Fm\overline{3}m$ space group with A2-type disorder yields a lattice parameter of  $a$$_{exp}$= 5.89$\pm$0.03 \AA. The same data was also fitted using the $Im\overline{3}m$ space group [Figure \ref{structure}(d)], yielding a lattice parameter of $a$$_{exp}$= 2.95$\pm$0.05 \AA. Here it is important to note that, the face-centered cubic (FCC) structure ($Fm\overline{3}m$) transforms into a body-centered cubic (BCC) structure ($Im\overline{3}m$) upon complete site mixing between Cr and Al atoms. Hence, the A2-disordered FCC structure ($Fm\overline{3}m$) with lattice parameter `a' is equivalent to the BCC structure ($Im\overline{3}m$) with lattice parameter `a/2' (see insets of Figure \ref{structure}(c) and (d)). The refined occupancies of Cr and Al atoms at various Wyckoff positions for both $Fm\overline{3}m$ and $Im\overline{3}m$ space groups are given in supplementary information (SI). Thus, Rietveld refinement confirms that both single-crystalline and polycrystalline Cr$_3$Al have complete site-mixing of Cr and Al atoms. In other words, this is characterised as A2-type disorder in Heusler structure having primitive BCC unit cell with $Im\overline{3}m$ space group. This complete disorder arises from the comparable atomic radii (R) and electronegativities ($\chi$) of Cr and Al atoms (R$_{Cr}$=1.40 Å, R$_{Al}$=1.25 Å;  $\chi_{Cr}=$1.66 and $\chi_{Al}=$1.61), a factor known to promote chemical disorder.  Such disorder is known to significantly influence the electronic, magnetic, and transport properties of Heusler alloys and can occur in both polycrystalline and thin-film forms \cite{Bainsla2017}.

 \subsection{Magnetization measurements}

 Figure \ref{magnetization}(a) and (c) show the magnetic susceptibility ($\chi$) vs. temperature ($T$) for polycrystalline and single crystalline samples respectively under 1000 Oe applied magnetic field, clearly indicating an anomaly at/around 800 K. The field dependent magnetization for the polycrystalline sample, as shown in Figure \ref{magnetization}(b), indicates a linear behavior with a very small hysteresis in between -500 Oe and +500 Oe. This confirms the possibility of an antiferromagnetic or weak ferrimagnetic ordering below 800 K. However, the large hysterisis observed in the ZFC-FC $M$-$T$ data for single crystal and polycrystal rules out the possibility of Neel antiferromagnetic ordering reported in previous thin film studies \cite{Boekelheide2012Mar,Boekelheide2012Aug,Zhao2019,TOYOKI2021}. Since the hysterisis is observed at a strong field of 1000 Oe, the existence of a short range ordering in the system is less probable for the whole temperature range below 800 K. Figure \ref{magnetization}(d) shows the $M$-$H$ data for single crystal sample, depicting a non linear behaviour in the field range of -3 kOe to +3 kOe without saturation and a coercive field of 200 Oe, reinforcing the soft ferrimagnetic nature of the system. A nearly compensated ordered moment lying in the range of 0.008 µ$_B$/ f.u is clearly evident for both single crystalline and polycrystalline samples.
 
  For ordered full Heusler alloys, the Slater-Pauling (SP) rule \cite{Slater1936,Pauling1938} relates the total magnetic moment ($M_{\mathit{t}}$) to the total number of valence electrons ($N_\text{v}$)  via the following expression:
\begin{equation}
M_\text{t} = (N_\text{v} - 24) \mu_B/\text{f.u.}
\end{equation}
For Cr\textsubscript{3}Al, the total number of valence electrons is 21, which, according to the SP rule, corresponds to a total magnetic moment of 3 $\mu_B$/f.u. However, the field dependent magnetization, as shown in Figure \ref{magnetization}(b) and (d), reveals an almost vanishing net magnetization.  This stark discrepancy of ordered moment is attributed to the presence of A2-disorder. Hence this behavior of temperature and field dependent magnetization is indicative of a fully compensated ferrimagnetic state. XMCD and temperature-dependent neutron diffraction measurements, discussed in the subsequent sections, further confirm the presence of long-range ferrimagnetic ordering with a compensated ordered moment.

 \subsection{X-ray magnetic dichroism (XMCD)}

 \begin
  {figure}[H]
	\centering
	\hspace*{-0.6cm}\includegraphics[width=1.23\linewidth]{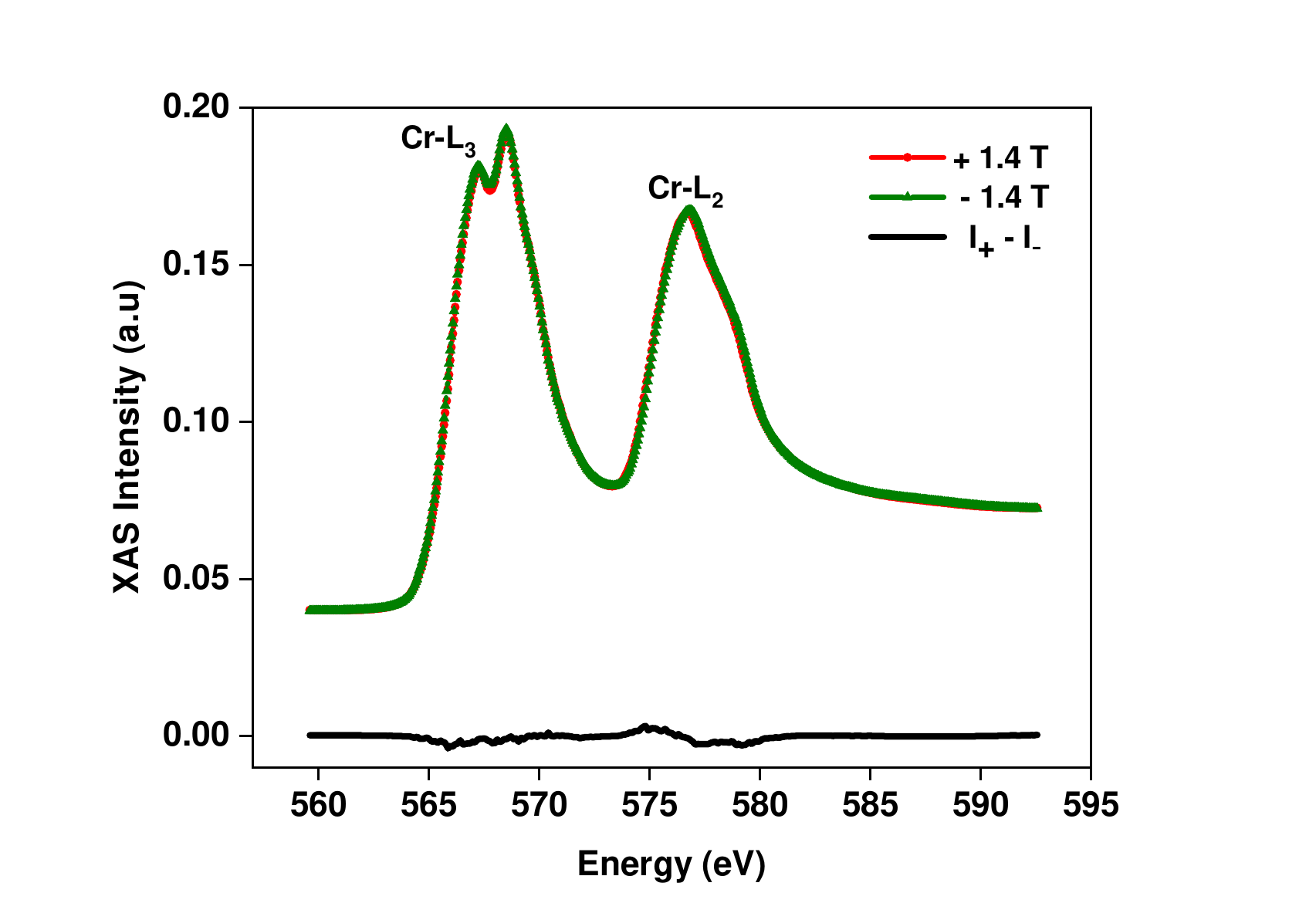}
	\caption{\justifying (Top) XAS measured with left circularly polarized X-rays under opposite directions of magnetic field and (bottom) the corresponding XMCD at the Cr L$_2$ and L$_3$ edges (black line). }
	\label{xmcd}
\end{figure}

The XMCD spectrum reflects the dichroic signal corresponding to difference in XAS spectra measured with left circularly polarized X-rays under opposite directions of the applied magnetic field.  Figure \ref{xmcd} shows the XAS spectra taken with left circularly polarized X-rays by changing the directions of the magnetic field and the corresponding XMCD signal at the Chromium L$_2$ and L$_3$ absorption edges. The XMCD signal intensity at zero reveals a net zero total magnetic moment on Cr atoms. This clearly rules out any ferromagnetic ordering in  Cr$_3$Al, rather points to an antiferromagnetic or fully compensated ferrimagnetic ordering. Hence the XMCD results are in complete agreement with our magnetization data. The observed peak splitting at the Cr-L$_3$ edge and the weak shoulder feature at the Cr-L$_2$ edge can be ascribed to the presence of inequivalent Cr sites experiencing distinct local crystal field environments. The XMCD spectrum evidently supports the observation of zero net magnetization in Cr$_3$Al.

\subsection{Neutron diffraction}
\begin{figure}[H]
	\centering
	\hspace*{-0.8cm}\includegraphics[width=1.15\linewidth]{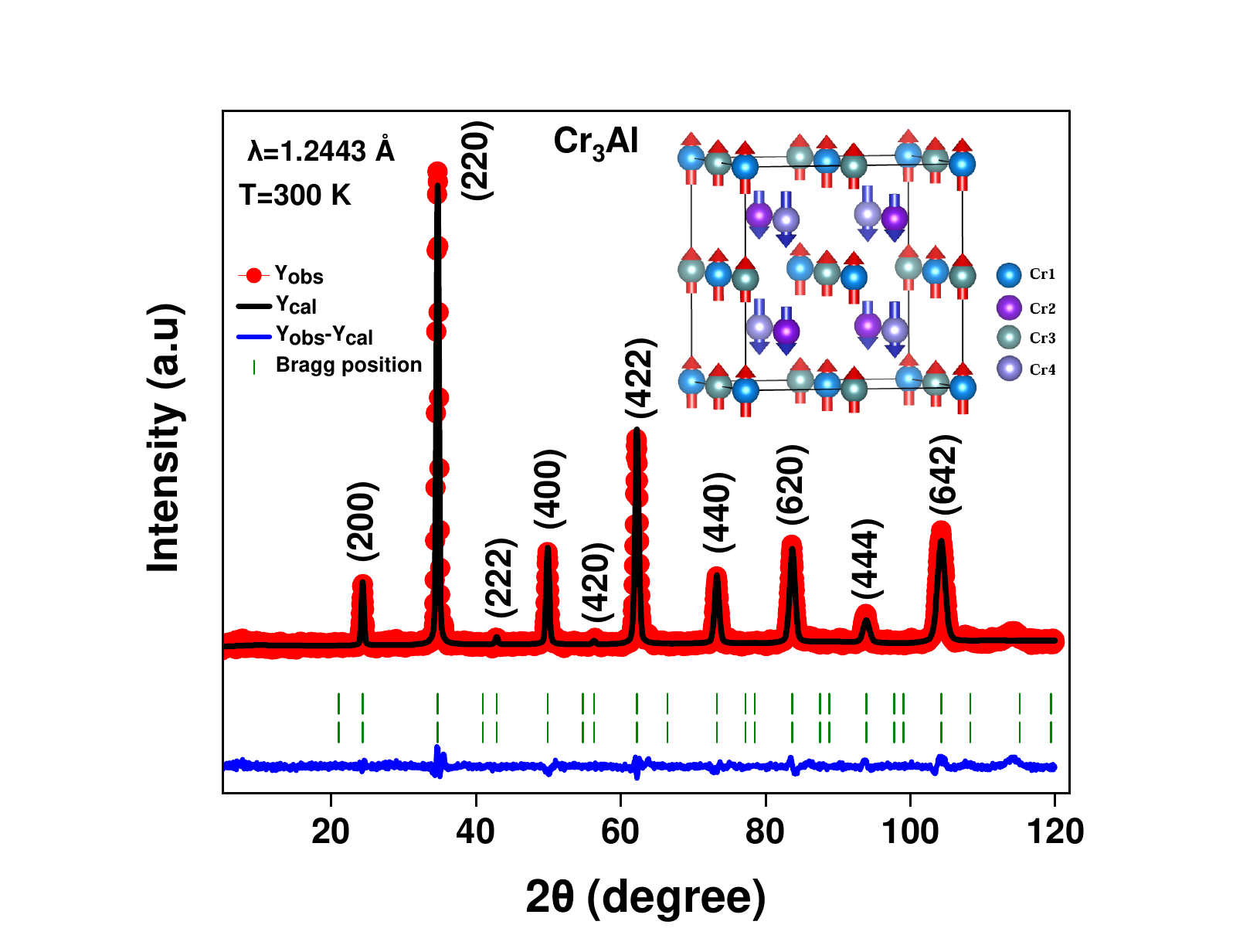}
	\caption{\justifying Observed neutron powder pattern of Cr$_3$Al at 300 K with the Rietveld refinement fit of the nuclear and magnetic structures. The green markers of the Bragg reflections for the nuclear (top lines) and magnetic reflections (bottom lines), along with the difference curve are shown at the bottom. The inset shows the alignment of the Cr1(0, 0, 0), Cr2(0.25, 0.25, 0.25), Cr3(0.5, 0.5, 0.5), and Cr4 (0.75, 0.75, 0.75) magnetic moments along the crystallographic axis. }
	\label{fig:sidebyside}
\end{figure}
Figure \ref{fig:sidebyside} shows the neutron powder diffraction pattern obtained at 300 K. The structure was modeled in the cubic $Fm\bar{3}m$ space group with A2-type disorder and Rietveld refinement yielded a lattice constant of \textit{a} = 5.89 $\pm$ 0.02 \AA. At first glance, we clearly observe that  the NPD data shows good agreement with the A2-type  disorder observed from Rietveld refinement of synchrotron XRD data. Interestingly, the (200) peak is highly prominent in the NPD data. Given the absence of the (200) reflection in the high-resolution synchrotron XRD data, and typical temperature evolution of the (200) peak in the NPD data upto 800 K (discussed below), it can be confirmed that it is purely magnetic in-origin. 
 
 To determine the magnetic structure, the 300 K diffraction pattern was analyzed by incorporating the magnetic contributions and using the lattice structure obtained from the structural refinement alone. The magnetic moment configuration was generated using the BasIREPS module of the FullProf program \cite{Reitveld}, based on the space group $Fm\bar{3}m$ and assuming ferrimagnetic (FIM) ordering. In the A2-disordered $Fm\bar{3}m$  structure, the two Cr sites 8$c$ (0.25, 0.25, 0.25), 4$a$ (0, 0, 0) and one Al site 4$b$ (0.5, 0.5, 0.5) is randomly occupied by Cr with certain probability. Hence the system comprises of magnetic moments originating from four distinct sites - Cr1, Cr2, Cr3 and Cr4 corresponding to the Wyckoff positions (0, 0, 0), (0.25, 0.25, 0.25), (0.5, 0.5, 0.5), and (0.75, 0.75, 0.75), respectively. For the FIM state, the magnetic propagation vector $\mathbf{k} = (0,0,0)$ was identified as the most appropriate using the k-search routine in FullProf and interestingly, the basis function obtained from this method reveals the moments are along c-axis of FCC unitcell.

   For the magnetic refinement of the diffraction pattern at 300 K, the parameters obtained from the structural refinement were used as the initial input and the $F\bar{1}$ magnetic space group was considered, with Cr1 and Cr3 moments aligned antiparallel to those of Cr2 and Cr4. Subsequently, the magnetic moments of the atoms were refined to obtain the microscopic magnetic moment values at each site. The antiparallel spin configuration of the neighboring nonequivalent Cr atoms resulted an almost compensated ordered moment of 0.1(1) $\mu_B$/f.u as shown in Table \ref{nd}, confirming the fully compensated ferrimagnetic behaviour, as also observed from magnetization and XMCD measurements. Refinement at 300 K yielded $\chi^{2} = 8.7$, $R_{B} = 1.95$, $R_{F} = 1.22$, and $R_{\mathrm{mag}} = 13.08$, suggesting reliable fits. The magnetic structure is shown in the inset of Figure \ref{fig:sidebyside}.
\begin{table}[H]
\centering
\caption{\justifying Structural and magnetic parameters obtained from the Rietveld 
refinement of the NPD pattern at 300 K.}
\label{nd}

\small
\setlength{\tabcolsep}{2.8pt}
\renewcommand{\arraystretch}{1.06}

\scalebox{0.8}{
\begin{tabular}{lccccc}
\hline
Atoms & X & Y & Z & Occupancy & $m$ ($\mu_B$) \\
\hline
Cr1 & 0.0000 & 0.0000 & 0.0000 & 1 & 0.855 $\pm$ 0.162 \\
Cr2 & 0.2500 & 0.2500 & 0.2500 & 1 & -0.921 $\pm$ 0.123 \\
Cr3 & 0.5000 & 0.5000 & 0.5000 & 1 & 0.764 $\pm$ 0.161 \\
Cr4 & 0.7500 & 0.7500 & 0.7500 & 1 & -0.800 $\pm$ 0.121 \\
\hline
\multicolumn{2}{l}{\textbf{Space group}} 
    & \multicolumn{4}{l}{$Fm\bar{3}m$} \\
\multicolumn{2}{l}{\textbf{Lattice parameters (\AA)}} 
    & \multicolumn{4}{l}{$a=b=c=5.89 \pm 0.02$} \\
\multicolumn{2}{l}{}
    & \multicolumn{4}{l}{$\alpha=\beta=\gamma=90^\circ$} \\
\multicolumn{2}{l}{\textbf{RF factor}} 
    & \multicolumn{4}{l}{1.22} \\
\multicolumn{2}{l}{\textbf{Bragg $R$ factor}} 
    & \multicolumn{4}{l}{1.95} \\
\multicolumn{2}{l}{\textbf{Magnetic space group}} 
    & \multicolumn{4}{l}{$F\bar{1}$} \\
\multicolumn{2}{l}{\textbf{Ordered moment}} 
    & \multicolumn{4}{l}{0.1(1) $\mu_B$/f.u.} \\
\hline
\end{tabular}}
\end{table}

\begin{figure}[H]
	\centering
	\hspace*{-0.7cm}\includegraphics[width=1.25\linewidth]{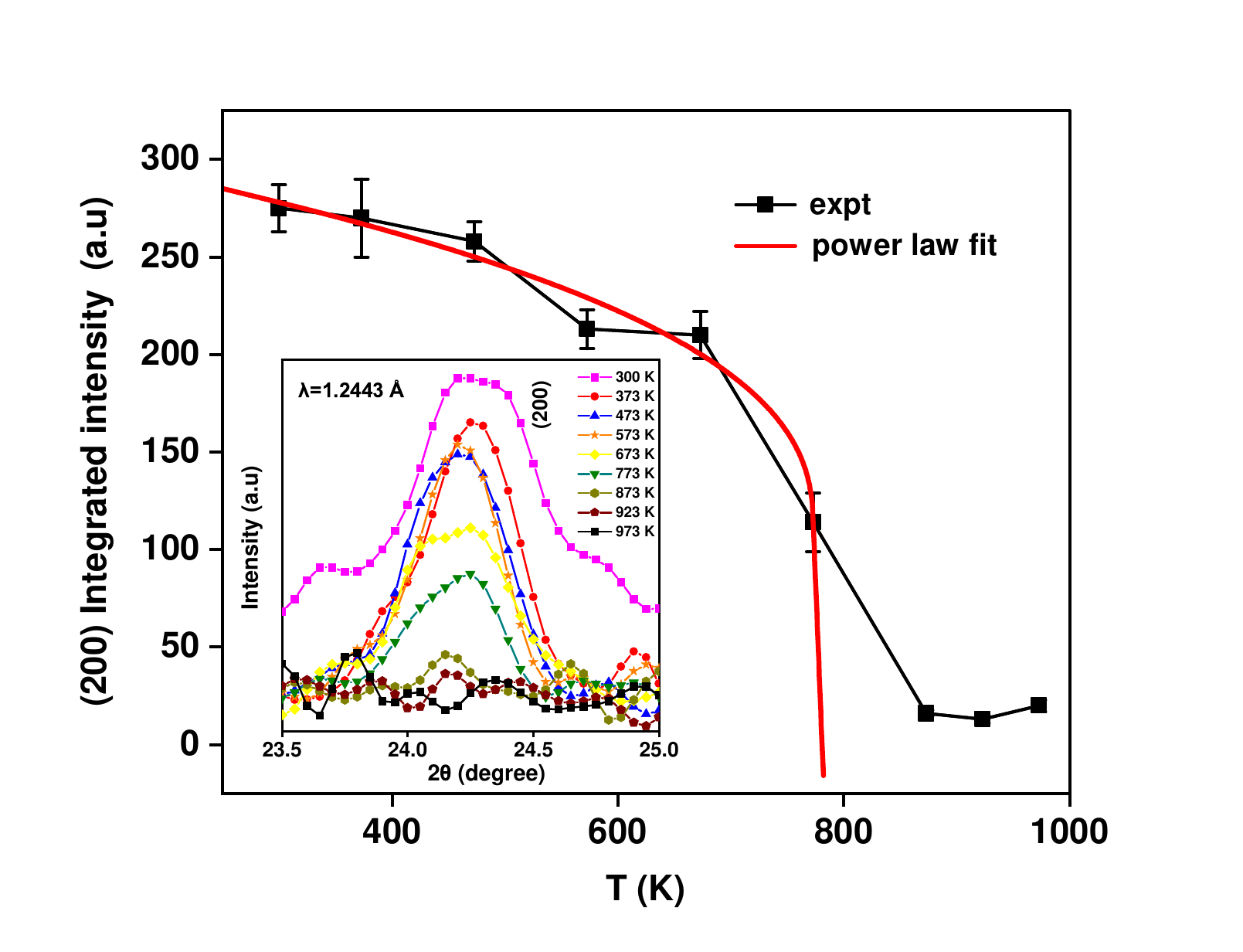}
	\caption{\justifying  Integrated intensity of the (200) magnetic reflection
 of Cr$_3$Al measured as a function of temperature.The solid red curve represents a power-law fit to the data with critical exponent $\beta$. The bottom left inset shows the intensity of (200) magnetic reflection versus temperature collected at 300, 373, 473, 573, 673, 773, 873, 923 and 973 K. The solid line is a guide to the eye.}
	\label{intg intensity}
\end{figure}

 The temperature-dependent measurements reveal that the intensity of the (200) peak decreases with increasing temperature and nearly vanishes above 773 K as depicted in the inset of Figure \ref{intg intensity}. The full NPD patterns at T= 300 K, 373 K, 473 K, 573 K, 673 K, 773 K, 873 K, 923 K and 973 K are shown in the supplementary information (SI). The integrated intensities of (200) peak at different temperatures were fitted with the power law, $I(T) = I_{0}\left(1 - \frac{T}{T_{C}}\right)^{2\beta}$. The fitting yielded a magnetic transition temperature of $T_{C} = 773 \pm 2$ K and a critical exponent of $\beta = 0.20 \pm 0.04$. The $\beta$ value is consistent with that reported for Cr$_2$CoAl \cite{gupt2023}.

  \subsection{Electrical transport and Hall measurements}

\begin{figure}[H]
	\centering
	\hspace*{-0.8cm}\includegraphics[width=1.135\linewidth]{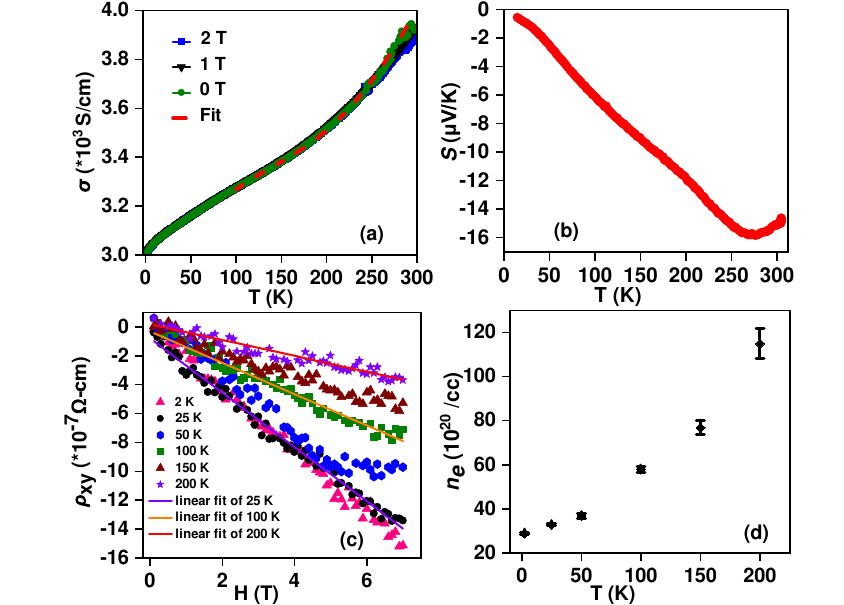}
	\caption{\justifying(a) Temperature dependent conductivity of Cr$_3$Al polycrystal at few magnetic fields with conductivity at zero field fitted with two-carrier model (b) The temperature variation of Seebeck coefficient,
$S$ in the range of 15 to 305 K (c) The field dependent Hall resistivity from 0 T to 7 T at few temperatures and the linear fit for the data at 25 K, 100 K and 200 K for calculating Hall coefficient (d) The temperature dependent carrier concentration for a temperature range from 2 K to 200 K.}
	\label{transport}
\end{figure}

Figure \ref{transport}(a) shows the $T$-dependence of longitudinal conductivity $\sigma_{xx}$ at 0, 1 and 2 T fields. The data shows a very weak $T$-dependence, temperature coefficient of conductivity d$\sigma$/d$T$ $>$0, hinting towards a semiconducting behaviour, unlike the thin film report of Cr$_3$Al  by Toyoki \textit{et al.} \cite{TOYOKI2021}. Also the effect of magnetic field on conductivity is almost negligible.
It is well known that for intrinsic semiconductors, the carrier concentration has exponential dependence on temperature. But the absence of such dependency makes it evident that there is a gapless semiconducting behaviour in the system. It can be a simple gapless semiconductor, spin gapless semiconductor or spin semimetal. To further investigate the electronic ground state, we have considered the two-carrier model for the $T$-dependent $\sigma$($T$) at zero field, where the entire transport mechanism is contributed from both electrons and holes \cite{Jamer2017, book}. As per this model, total conductivity can be written as:

\begin{equation}
\sigma(T) = \sigma_e + \sigma_h = e n_e \mu_e + e n_h \mu_h
\end{equation}

where $\sigma_e$ and $\sigma_h$ represents the contribution from electrons and holes. The thermally created carrier concentration of respective carriers, $n_e$ and $n_h$ can be expressed as \[
n_e \sim n_{e0} \exp\left( \frac{-\Delta E_e}{k_B T} \right), \quad
n_h \sim n_{h0} \exp\left( \frac{-\Delta E_h}{k_B T} \right)
\] where $\Delta E_e$ and $\Delta E_h$ are the gaps for the two carriers. The mobilities are given by\[\mu_i = \left( a_i T + b_i \right)^{-1} = \frac{\mu_{i0}}{a'_i T + 1}
(i=e,h)\] The values of $a$ and $b$ coefficients are different for both elctrons and holes. The term containing $a$ represents the contribution to mobility due to phonon scattering, whereas the second term containing $b$ is independent of temperature and it involves the mobility due to defect scattering at $T$=0 K. Introducing these expressions we obtain the final form of $\sigma$ as, 
\begin{equation}
\sigma(T) = A_e(T) \, e^{-\Delta E_e / k_B T} + A_h(T) \, e^{-\Delta E_h / k_B T}
\end{equation}
were $A_i(T) = \dfrac{en_{i0}\mu_{i0}}{1 + a_i'T}$ for $i = e$ and $h$. The measured $\sigma$($T$) data fits very well with this modified activated transport model having a non-exponential dependence within the temperature range of 100 K to 300 K, as shown in Figure \ref{transport}(a). The fitting parameters $a'_e$ and $a'_h$ are (0.10 $\pm$ 0.03)$\times$10$^{-3}$ and (0.10 $\pm$ 0.05)$\times$10$^{-2}$ respectively. The values are close to zero which implies the fact that mobility of Cr$_3$Al is substantially ruled by defect scattering as compared to that from phonons. The energy gaps extracted from the fit are 0.92 $\pm$ 0.07 meV for electrons and 94 $\pm$ 4 meV for holes. Hall effect measurements indicate that electrons constitute the majority carriers, implying that the observed lower-energy gap can be attributed to that. At $T$=300 K, $\sigma_e$ and $\sigma_h$ are estimated to be 3440 $\pm$ 30 S/cm and 570 $\pm$ 90 S/cm respectively. The electron conductivity is much larger than hole conductivity. The fitted parameters $a'_e$, $a'_h$, $\Delta E_e$, and $\Delta E_h$ were found to lie within the same range as those reported for other SGS systems, such as CrVTiAl \cite{Venkateswara2018} and Co$_{1+x}$Fe$_{1-x}$CrGa ($x$ = 0.1 to 0.4) \cite{Rani2019} [further comparison of different properties with other SGS materials is tabulated in Table \ref{sgs_comparison}]. The calculated mobility of electrons is found to be $\sim$ 1.8 cm$^2$/(V.s)  which is typically low compared to conventional semiconductors. The relatively low carrier mobility observed in Cr$_3$Al can be ascribed to the chemical disorder, which enhances electron scattering in the system \cite{Venkateswara2018}.

The field dependence of Hall resistivity for the polycrystalline Cr$_3$Al at few temperatures between 2 K and 200 K is given in Figure \ref{transport}(c). The Hall resistivity exhibits a linear behavior with a negative slope, indicating a negative Hall coefficient and confirming electrons as the dominant charge carriers. The linear behaviour indicates the absence of any anomalous Hall contribution in the system. The carrier concentration ($n$$_e$) is calculated from the Hall data at constant temperatures, as shown in Figure \ref{transport}(d). $n$$_e$ increases from 29$\times$10$^{20}$/cc to 115$\times$10$^{20}$/cc over the temperature range of 2 K to 200 K. The calculated values are of the same order as that reported for the other fully compensated ferrimagnetic spin gapless semiconductors (FCF SGS) CrVTiAl \cite{Venkateswara2018}, but quite higher than the ferromagnetic SGS Mn$_2$CoAl ($10^{17}$/$cm^3$) \cite{44_Ouardi2013}. Moreover, disorder can substantially modify the electronic structure by introducing localized states in the vicinity of the Fermi level, which act as thermally activated carrier reservoirs. Consequently, these defect-induced states account for the $T$-dependent enhancement in carrier density as observed in Cr$_3$Al, in contrast to the nearly temperature-independent behavior reported for other SGS systems. Our spin-polarized density functional theory (DFT) calculations further corroborate the SGS behavior observed in this system.

  \subsection{Thermal transport}
  
  The Seebeck coefficient ($S$) provides critical insights into the type and density of charge carriers and hence is an important quantity to assess the nature of electronic band structure. For the SGS which clearly acquire a very unique band structure with a zero gap for one spin band and a finite gap for the other, the thermopower response will be highly sensitive to carrier asymmetry between electrons and holes. Thus a nearly zero Seebeck coefficient manifesting simultaneous presence of both electron and hole carriers is a key signature of gapless and spin polarized band structure. The total Seebeck coefficient with contribution from both electrons and holes can be written as:
\begin{equation}
S = \frac{S_e \sigma_e + S_h \sigma_h}{\sigma_e + \sigma_h}
\end{equation}
where $S_e<0$ and $S_h>0$ and $\sigma_e$, $\sigma_h$ are electron and hole conductivities. Figure \ref{transport}(b) shows the temperature dependence of Seebeck coefficient measured in the $T$-range of 15 K to 305 K.

The S value varies in between -0.5 to -15 $\mu$V/K for a temperature range of 15 to 305 K. The consistently negative Seebeck coefficient across the entire $T$-range further supports the Hall data, indicating that electrons are the majority charge carriers.  The value remains significantly lower compared to that of conventional semiconductors, having Seebeck coefficients in the range of 100 to 1000 $\mu$V/K. The appreciably low value of Seebeck coefficient (\(|S|\)$<$20 $\mu$V/K ) suggests electron and hole compensation, as observed in other SGS systems like Mn$_2$CoAl \cite{44_Ouardi2013}, Co$_{1+x}$Fe$_{1-x}$CrGa ($x$ = 0.1 to 0.4) \cite{Rani2019}. These aspects support the SGS behaviour of Cr$_3$Al. However, in comparison with the reported SGS systems, $S$ is slightly higher and has a non trivial dependence on temperature for Cr$_3$Al. This can be attributed to atomic disorder in the system, which possibly creates minor additional states at/near the Fermi energy and enhances carrier conduction unlike the other reported SGSs. 

 \section{Theoretical Results}

 To start with, we performed ab-initio calculations of the ordered DO$_3$ structure of Cr$_3$Al, where the most stable configuration places Cr$_1$, Cr$_2$, Cr$_3$ and Al at the 4$b$, 4$c$, 4$d$, and 4$a$ Wyckoff positions, respectively. Cr$_2$ and Cr$_3$ are structurally equivalent due to identical local environments. Starting from a ferromagnetic (FM) configuration, the structure was fully relaxed, yielding a lattice parameter of 5.958 \r{A}. Post-relaxation, Cr$_2$, Cr$_3$ flip their spins and couple antiferromagnetically with Cr$_1$ resulting in a finite net magnetic moment of -2.83 $\mu_{B}$ (Cr$_2$, Cr$_3$) and 2.72 $\mu_{B}$ (Cr$_1$), while Al remains nearly nonmagnetic (0.036 $\mu_{B}$), giving a total unit cell moment of $\vert\mu_{tot}\vert$= 2.91 $\mu_{B}$/f.u., consistent with previous literature reports \cite{Gao2013,Li2009}.
 \begin{figure}[H]
    \centering
    \hspace*{-0.375cm}\includegraphics[width=1.18\linewidth]{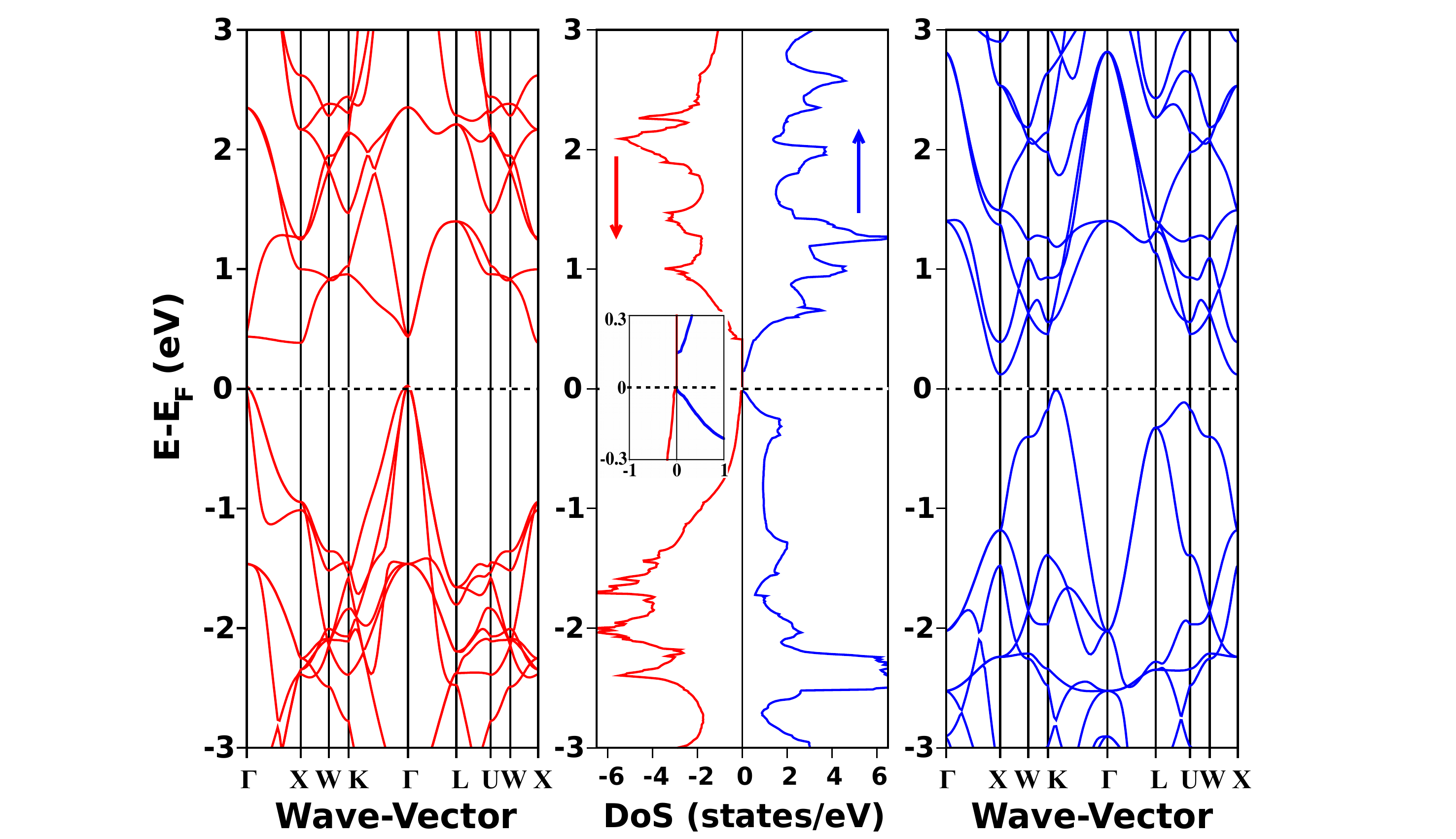}
    \caption{Spin polarized density of states (DoS) and electronic band structures (left panel: spin down channel and right panel: spin up channel) for the DO\textsubscript{3} ordered structure of Cr\textsubscript{3}Al. The inset shows a zoomed-in view of the DoS near the Fermi level (E\textsubscript{F}).}
    \label{figure-7}
\end{figure}
Figure \ref{figure-7} shows the spin polarized density of states (DoS) and electronic band structure for the DO$_3$ ordered structure indicating a magnetic semiconducting nature with a band gap (E\textsubscript{g}) of 0.13 eV and 0.40 eV for spin up and down channels respectively, as evident from the inset of Figure \ref{figure-7}. It should be noted that the value of E\textsubscript{g} for spin up channel is quite small and the system can become an SGS if this value approaches zero. The presence of valence band maximum and conduction band minimum at different high symmetry point of the Brillouin zone (BZ) is suggestive of an indirect band gap nature of the ordered Cr\textsubscript{3}Al. 
It is important to note that the present calculations are done using the meta-GGA-based functional which captures the correlation effects better. Consequently, our results contrast with earlier GGA-based predictions of metallic or half-metallic behavior. This difference stems from the GGA functional's limitations in treating the strongly correlated Cr d-electrons. The meta-GGA functional better captures exchange-correlation effects, especially for systems with on-site interactions and inequivalent atoms, due to the fact that it takes into account the local kinetic energy density corrections which accurately treats different chemical bonds compared to the conventional GGA functional. The meta-GGA functional also satisfies all the 17 exact mathematical constraints that a semi-local functional must satisfy, highlighting its mathematical robustness. In contrast, the GGA functional satisfies only 11 out of the exact 17 constraints. 

\begin{figure}[H]
    \centering
    \hspace*{-0.7cm}\includegraphics[width=1.2\linewidth]{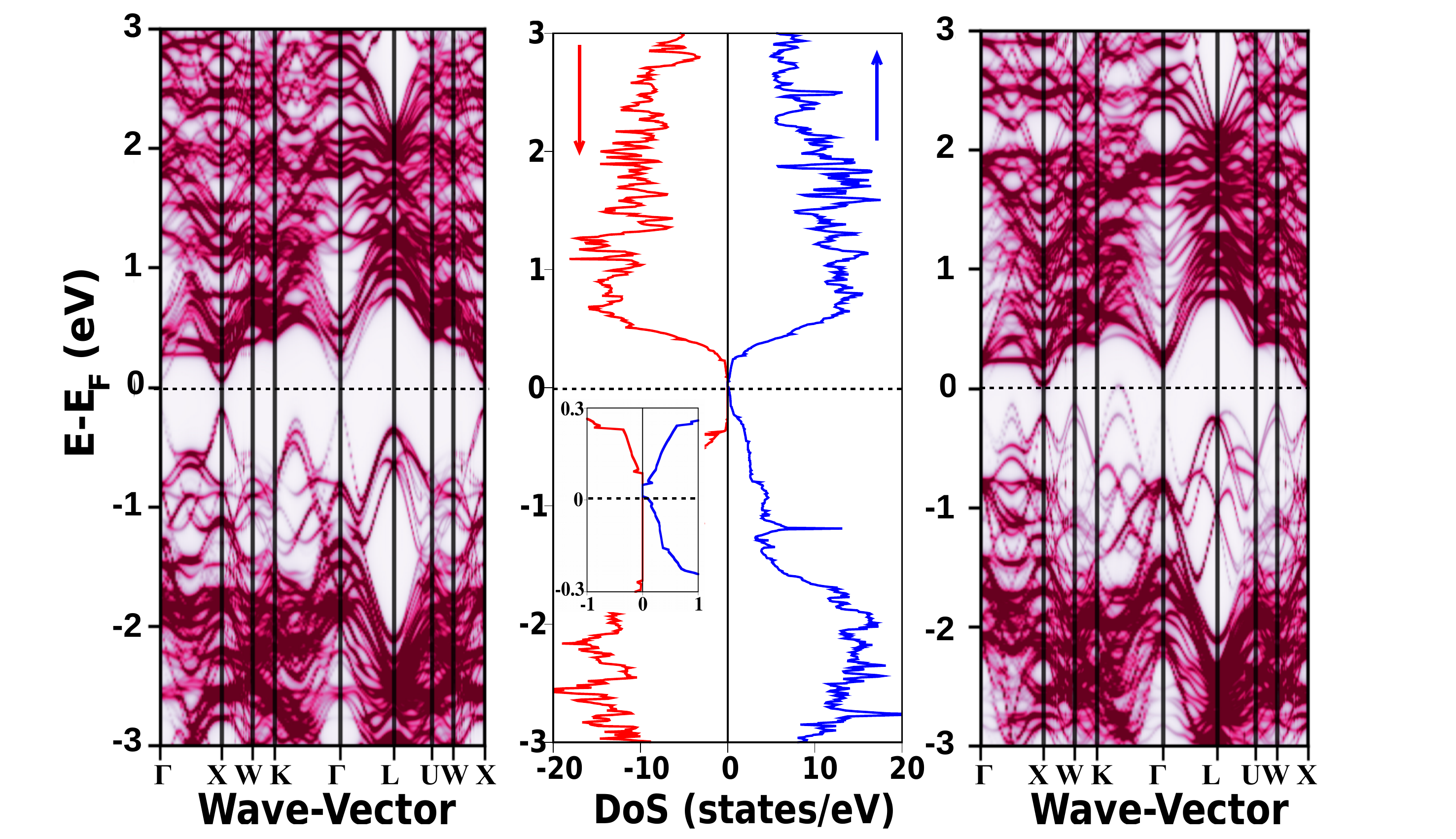}
    \caption{Spin polarized density of states (DoS) and unfolded electronic band structures for the A2-disordered structure of Cr\textsubscript{3}Al. The inset shows a zoomed-in view of the DoS near the Fermi level (E\textsubscript{F}).}
    \label{figure-8}
\end{figure}

 The XRD analysis (see Figure \ref{structure}) however suggests the presence of an A2-disorder. To capture the disorder effect, we have generated a 16-atom SQS structure. The theoretically optimized lattice parameter of the SQS structure is 5.937 \AA\ which is in close agreement with our measured value. Interestingly, the net magnetic moment vanishes with a negligibly small value of 0.0072 $\mu_B$/f.u. The substitutional disorder disrupts the exchange interactions and weakens the ferromagnetic coupling. Due to the change in the local atomic environment, the magnetic moments of several Cr atoms in the SQS cell align antiferromagnetically while Al atoms remain non-magnetic. The site-projected magnetic moments of the SQS cell are provided in Table \ref{sqs} of supplementary information (SI). These are the moments of Cr$_i$ atoms located at different random sites (4a, 4b, 4c and 4d). However, if we calculate the average moments at 4 different Wyckoff sites, it turns out to lie in the range 1.1 to 1.3 $\mu_B$, with +ve moments at 4a and 4c sites while -ve moments at 4b and 4d sites, eventually giving a fully compensated net moments. These averaged site projected moments compare well with those measured using NPD data, shown in Table 2. Figure \ref{figure-8} shows the spin polarized DoS and the unfolded electronic band structure for this SQS structure of Cr\textsubscript{3}Al. The band gap for both the spin channels reduces to a value of 0.35 eV (for spin-down) and 0.01 eV (for spin-up). Notice that, the band gap for spin up channel is sufficiently small to qualify it as a SGS candidate. It should be noted that for the spin up channel, a very small density of states appear at the Fermi level (see inset of Figure \ref{figure-8}) which is due to an off-symmetry crossing of valence band. The closing of gap by the conduction band at X-point for spin up channel and a modest gap in the spin down channel are the prerequisite revealing the SGS nature of the disordered Cr\textsubscript{3}Al.

\begin{figure}[H]
    \centering
    \hspace*{-0.2cm}\includegraphics[width=1.055\linewidth]{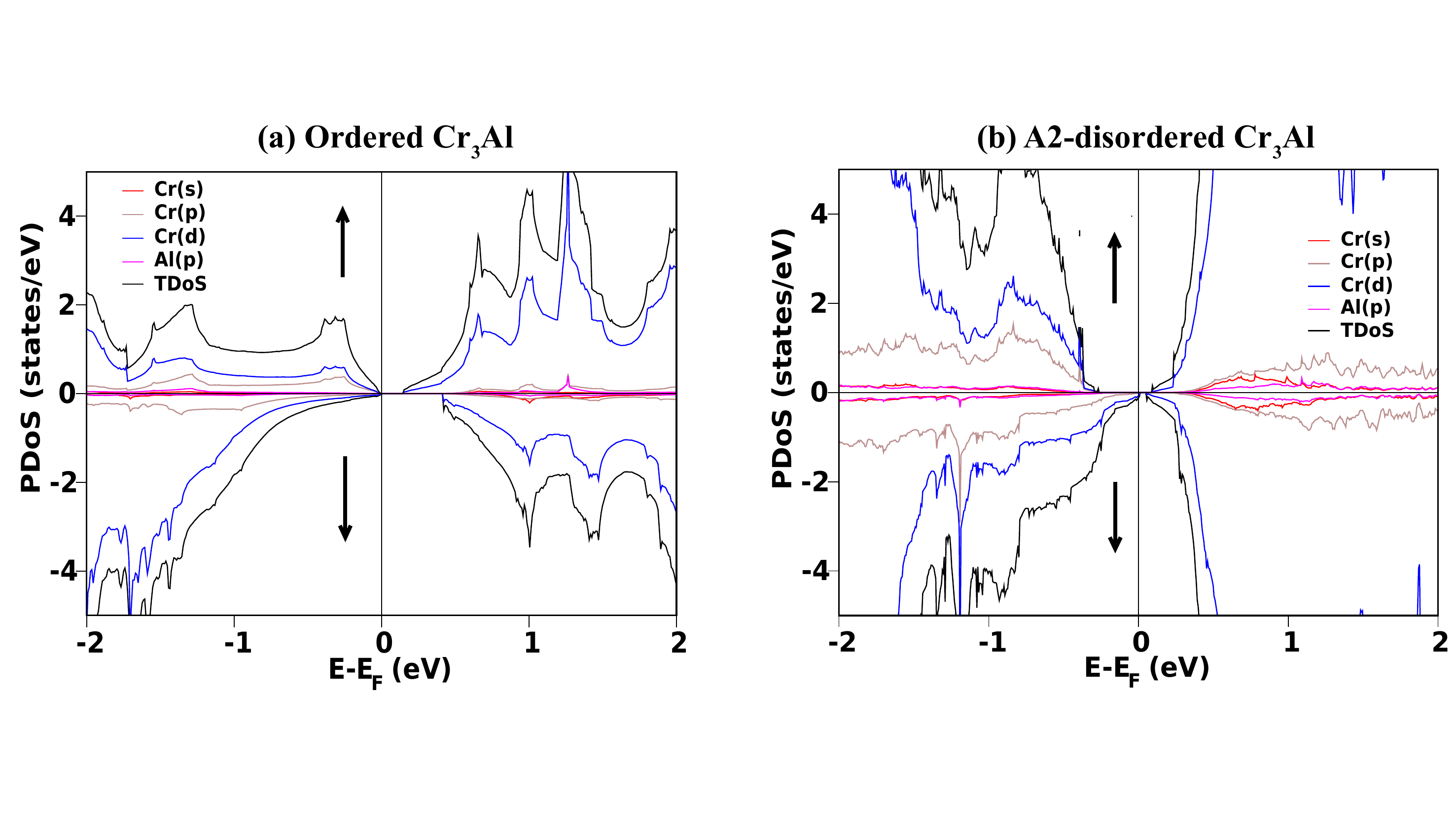}
    \caption{Spin polarized partial density of states for (a) DO\textsubscript{3} ordered  and (b) A2-disordered structure of Cr\textsubscript{3}Al.}
    \label{figure-9}
\end{figure}

 The dominant orbital contributions to the total DoS for the ordered- and disordered-structures are shown in Figure \ref{figure-9}. For both the structures, the states near the Fermi level are dominated by the Cr(d) orbitals with very small traces of Cr(p) and Cr(s) in the ordered structure. However, for the disordered structure, the contribution from the Cr(p) becomes non-negligible. The Al(p) manifold remains very marginal and hence it is less likely to be involved in dictating the magnetic and transport properties of Cr$_3$Al.

\begin{table*}[h!]
\caption{Comparison of electrical, thermal, and magnetic parameters of Cr$_3$Al with previously reported spin-gapless semiconductor (SGS) systems.}
\label{sgs_comparison}
\centering

\small
\setlength{\tabcolsep}{2pt}
\renewcommand{\arraystretch}{1.6}

\resizebox{1.015\textwidth}{!}{%
\begin{tabular}{lccccc}
\toprule
\textbf{SGS systems / Parameters} 
& Mn$_2$CoAl \cite{44_Ouardi2013, Xu2014} 
& CrVTiAl \cite{Venkateswara2018} 
& Co$_{1+x}$Fe$_{1-x}$CrGa \cite{Rani2019, Mishra2023} 
& CoFeMnSi \cite{Bainsla2015, huarui2019} 
& Cr$_3$Al (This work) \\
\midrule

$\Delta E_g$ (eV) 
& 0.17 ($\downarrow$)
& 0.36 ($\uparrow$) 
& 0.12 ($\downarrow$) (x=0); 0.08 ($\downarrow$) (x=0.125); 0.10 ($\downarrow$) (x=0.25)
& 0.62 ($\downarrow$)  
& 0.35 ($\downarrow$) \\

N($E_F$) (states/eV) 
& -- 
& 0.08 ($\downarrow$) 
& 0.17 ($\uparrow$) (x=0); 0.30 ($\uparrow$) (x=0.125); 0.24 ($\uparrow$) (x=0.25)
& 0.28 ($\uparrow$) 
& 0.11 ($\uparrow$) \\

$E_g$ (electron) (meV) 
& -- 
& 63.4 
& 76.2 (x=0.1); 89.5 (x=0.2); 53.5 (x=0.3) 
& -- 
& 0.92 \\

$E_g$ (hole) (meV) 
& -- 
& 0.3 
& 0.1 (x=0.1, 0.2, 0.3) 
& -- 
& 94 \\

$\mu$ (cm$^2$/V·s) 
& 0.45 (Thin film at 5 K)
& 2.2 
& 44.37 (Thin film at 300 K, x=0)
& 46 (at 300 K)
& 1.8 (at 300 K) \\

$n$ (cm$^{-3}$) 
& 10$^{17}$ 
& 10$^{22}$ 
& 10$^{21}$ (Thin film, x=0)
& 10$^{19}$ 
& 10$^{22}$ \\

$\sigma_{xx}$ at 300 K (S/cm) 
& 2440 
& 3830 
& 2290–3294 (x=0.1 to 0.4) 
& 2980 
& 3925 \\

$S$ ($\mu$V/K) 
& Almost zero below 150 K; $\sim$1 at RT 
& -- 
& Nearly zero for 2–40 K; 1.1 for 40–300 K (x=0.1) 
& -- 
& –15 at RT \\

$T_C$ (K) 
& 720 
& 710 
& 690–870 (x=0 to 0.5) 
& 620 
& 800 \\

$M_\text{tot}$ ($\mu_B$/f.u.) 
& 2 
& 10$^{-3}$ 
& 2.1–2.5 (x=0.1 to 0.5) 
& 3.7 
& 0.008 \\
\bottomrule
\end{tabular}
} 

\end{table*}

\section{Discussion, Conclusion and Future Outlook}
The present work establishes Cr$_3$Al as a chemically disordered Heusler alloy that defies conventional expectations regarding the interplay between disorder, magnetism, and electronic transport. Although Heusler-based SGSs are typically realized in structurally ordered compounds, our combined experimental and theoretical investigation demonstrates that SGS behavior and FCF can not only coexist but remain remarkably robust within a fully A2-disordered environment. This highlights a broader and conceptually important outcome: chemical disorder, often considered detrimental to spin-polarized transport, can under certain circumstances stabilize functional electronic and magnetic states rather than suppress them.

Structurally, both single-crystalline and polycrystalline Cr$_3$Al adopt complete Cr/Al site mixing. Despite this extensive disorder, magnetometry, XMCD, and neutron diffraction collectively reveal a compensated ferrimagnetic ground state with an exceptionally small net moment (~10$^{-3}$$\mu_B$/ f.u.) and a high magnetic ordering temperature of 773 ± 2 K. The persistence of long-range ferrimagnetic ordering under full A2 disorder underscores the intrinsic stability of the magnetic sublattices. This behavior is also validated by first-principles calculations, where the SQS-based disordered model naturally yields an almost zero net moment through antiparallel Cr spin configurations across mixed sites.

Equally notable are the transport characteristics. The weak temperature dependence of electrical conductivity, low Seebeck coefficient, and compensated electron–hole transport signatures provide a coherent picture of SGS behavior. The conductivity ($\sigma_{xx}$) at 300 K is estimated to be 3925 S/cm, which is comparable to the values reported for other SGS systems (see Table 3). A low electron mobility value of 1.8 cm$^2$ V$^{-1}$ s$^{-1}$, obtained from the two-carrier conductivity fit, reflects strong carrier scattering associated with chemical disorder. Unlike conventional SGSs—where carrier densities remain nearly temperature-independent, Cr$_3$Al exhibits a pronounced temperature-driven increase in carrier concentration (increases from 29$\times$10$^{20}$/cc to 115$\times$10$^{20}$/cc for a range of 2 K to 200 K temperature). We attribute this to disorder-induced electronic states near the Fermi level, which act as thermally activated carrier reservoirs. This further suggests that disorder does not merely perturb SGS behavior but actively reshapes it, enabling transport regimes that differ fundamentally from those reported in ordered Heusler SGSs such as Mn$_2$CoAl and CrVTiAl.

Our theoretical calculations offer additional insights. In the ordered DO$_3$ structure, Cr$_3$Al exhibits magnetic semiconducting gaps in both spin channels, incompatible with the experimentally observed SGS features. In contrast, introduction of realistic A2 disorder through SQS modeling collapses the spin-up gap to near zero while retaining a finite spin-down gap—directly mirroring the transport signatures. The strong agreement between theory and experiment reinforces the conclusion that A2-type disorder is the key driving factor enabling SGS behavior in Cr$_3$Al.

Table~\ref{sgs_comparison} display a comparative summary of the electrical, thermal transport and magnetic parameters of Cr$_3$Al with other reported spin-gapless semiconductor systems Mn$_2$CoAl, CrVTiAl, CoFeMnSi, and Co$_{1+x}$Fe$_{1-x}$CrGa. In totality, Cr$_3$Al represents the first experimentally verified fully disordered Heusler compound that simultaneously exhibits SGS transport and fully compensated ferrimagnetism, with a Curie temperature approaching 800 K.  The key attributes found in disordered Cr$_3$Al— i.e. zero net magnetization, disorder-tolerant SGS features, and high-temperature stability— are thus pivotal for future spintronic devices where stray-field-free operation and strong spin selectivity are essential.

In terms of future outlook, the coexistence of SGS behavior and fully compensated ferrimagnetism in a fully A2-disordered Heusler alloy opens several promising research directions. Future efforts can explore controlled tuning of disorder—via chemical substitution, film growth techniques, or strain engineering—to systematically modulate band topology and spin polarization. The unusually robust high-temperature magnetic compensation further motivates the development of Cr$_3$Al-based thin films, heterostructures, and spin-filter devices where disorder engineering could serve as an additional design parameter. More broadly, Cr$_3$Al provides a compelling prototype for uncovering disorder-enabled functional states in intermetallics, suggesting that chemically disordered Heusler alloys may represent a rich, yet largely unexplored, materials platform for next-generation spintronic technologies.

\section*{Acknowledgments}
We thank Dr. Velaga Srihari, HP \& SRPD, BARC, Mumbai and Indus 2, BL-11 beamline for the Synchrotron XRD measurements, Shri Sudip Kumar Nath and Dr. Shreyashkar Dev Singh, APSUD, RRCAT, Indore for XMCD measurements and Prof. Ramesh Chandra Nath (IISER Trivandrum) for his initial support with the electrical transport measurement facility. We also thank Dr. S. M. Yusuf for initiating the collaboration on Neutron scattering. Reshna thanks the CIF, IIT Palakkad, and CIF, IISER Bhopal (magnetisation measurements) for the experimental facilities. Soham acknowledges the financial support from DST INSPIRE Faculty grant, Pooja Vyas acknowledges IIT Bombay for financial support through the Institute Post Doctoral Fellowship and S.K.K. acknowledges CSIR for research fellowship.
\clearpage

\section{Supplementary Information}
\subsection{Ordered and Disordered Structure of X$_3$Z Alloy}

\begin{figure}[H]
	\centering
	\hspace*{-0.6cm}\includegraphics[width=0.95\linewidth]{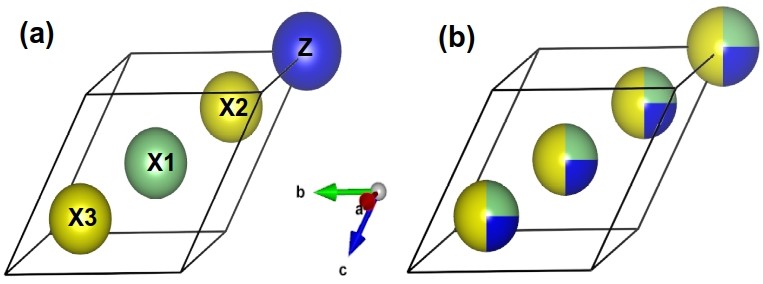}
	\caption{Primitive unit cell of (a) DO$_3$ ordered and (b) A2-disordered X\textsubscript{3}Z alloy.}
	\label{fi1}
\end{figure}

\subsection{DFT calculation details}

The ordered Cr\textsubscript{3}Al was first fully relaxed assigning a ferromagnetic configuration. However, the Cr\textsubscript{2} and Cr\textsubscript{3} atoms aligns antiferromagnetically with Cr\textsubscript{1} atom with an equilibrium lattice constant of 5.958 \AA\ 
. The site projected moments and the absolute net magnetization $\vert\mu_{tot}\vert$ of an ordered Cr\textsubscript{3}Al is given in Table~\ref{tab:atom_moments}.

\begin{table}[h!]
\centering
\scalebox{0.75}{   
\begin{tabular}{|c|c|c|c|}
\hline
\textbf{Site No.} & \textbf{Atom} & \textbf{Site Projected Moment ($\mu_{\mathrm{B}}$)} & \textbf{$|\mu_{\mathrm{tot}}|$ ($\mu_{\mathrm{B}}$/f.u.)} \\
\hline
1 & Cr$_1$ & 2.72 & \multirow{4}{*}{2.91} \\
2 & Cr$_2$ & -2.83 & \\
3 & Cr$_3$ & -2.83 & \\
4 & Al     & 0.036 & \\
\hline
\end{tabular}
}
\caption{Site projected moments and absolute net magnetization of ordered Cr\textsubscript{3}Al.}
\label{tab:atom_moments}
\end{table}

To simulate an A2-disordered cell, a 16-atom SQS was simulated using the mcsqs ATAT code \cite{VANDEWALLE2013}. The genearted structure was then fully relaxed with an initial ferromagnetic spin configuration. Post relaxation, several Cr atoms align antiferromagnetically due to the change in the local atomic environment. The Al atoms however, still remain nonmagnetic. The equilibrium lattice constant of the disordered structure is 5.937 \AA. The table~\ref{sqs} shows the site-projected magnetic moments and absolute net magnetization of an A2-disordered 16-atom cell.

\begin{table}[H]
\centering
\footnotesize                      
\setlength{\tabcolsep}{3pt}        
\renewcommand{\arraystretch}{0.9}  

\caption{Site projected magnetic moments and total magnetic moment per formula unit.}
\begin{tabular}{|c|c|c|c|}
\hline
\textbf{Site No.} & \textbf{Atom} & 
\textbf{Site Projected Moment ($\mu_{\mathrm{B}}$)} & 
\textbf{$|\mu_{\mathrm{tot}}|$ ($\mu_{\mathrm{B}}$/f.u.)} \\
\hline
1  & Cr$_1$ &  2.816  & \multirow{16}{*}{0.0072} \\
2  & Cr$_1$ &  2.816  & \\
3  & Cr$_1$ &  2.517  & \\
4  & Cr$_1$ &  2.517  & \\
5  & Cr$_2$ & -2.443  & \\
6  & Cr$_2$ & -2.443  & \\
7  & Cr$_2$ & -2.669  & \\
8  & Cr$_2$ & -2.669  & \\
9  & Cr$_3$ &  2.537  & \\
10 & Cr$_3$ &  2.537  & \\
11 & Cr$_3$ & -2.721  & \\
12 & Cr$_3$ & -2.721  & \\
13 & Al     & -0.049  & \\
14 & Al     &  0.026  & \\
15 & Al     &  0.026  & \\
16 & Al     & -0.049  & \\
\hline

\end{tabular}
\label{sqs}
\end{table}

\subsection{Synchrotron XRD refinement data of polycrystal}

\begin{table}[H]
\centering
\footnotesize        
\setlength{\tabcolsep}{4pt}      
\renewcommand{\arraystretch}{1.6} 

\caption{\label{tab:table3}Refined occupancies of Cr and Al atoms at various Wyckoff positions for FCC structure, R$_p$ = 20.7, R$_{wp}$ = 18.9, R$_{exp}$ = 16.84.}

\begin{tabular}{|l|c|c|c|}
\hline
\multicolumn{4}{|c|}{\textbf{Space group:} $Fm\overline{3}m$ (SG No.\ 225)} \\
\hline
\textbf{Atom} & \textbf{Wyckoff} & \textbf{(x, y, z)} & \textbf{Occ.} \\
\hline
Cr1 & 8$c$ & (0.25, 0.25, 0.25) & 0.03127 \\
Al1 & 8$c$ & (0.25, 0.25, 0.25) & 0.01040 \\
Cr2 & 4$b$ & (0.5, 0.5, 0.5) & 0.01560 \\
Al2 & 4$b$ & (0.5, 0.5, 0.5) & 0.00523 \\
Cr3 & 4$a$ & (0, 0, 0) & 0.01560 \\
Al3 & 4$a$ & (0, 0, 0) & 0.00523 \\
\hline
\end{tabular}
\end{table}

\begin{table}[H]
\centering
\footnotesize                 
\setlength{\tabcolsep}{4.5pt} 
\renewcommand{\arraystretch}{1.8} 

\caption{\label{tab:table3}Refined occupancies of Cr and Al atoms at various Wyckoff positions for BCC structure, R$_p$ = 31.1, R$_{wp}$ = 26.0, R$_{exp}$ = 16.88.}

\begin{tabular}{|l|c|c|c|}
\hline
\multicolumn{4}{|c|}{\textbf{Space group:} $Im\overline{3}m$ (SG No.\ 229)} \\
\hline
\textbf{Atom} & \textbf{Wyckoff} & \textbf{(x, y, z)} & \textbf{Occ.} \\
\hline
Cr1 & 2$a$ & (0, 0, 0) & 0.75072 \\
Al1 & 2$a$ & (0, 0, 0) & 0.25959 \\
\hline
\end{tabular}
\end{table}

The XRD Rietveld refinement was carried out in the cubic $Fm\overline{3}m$ space group, with A2-type disorder and $Im\overline{3}m$ space group using the \textsc{FullProf} Suite. This analysis yielded the aforementioned Cr and Al site occupancies at the respective Wyckoff positions. The refined data confirms the composition of Cr$_3$Al.

\subsection{Neutron diffraction}
\begin{figure}[H]
	\centering
	\hspace*{-0.6cm}\includegraphics[width=1.16\linewidth]{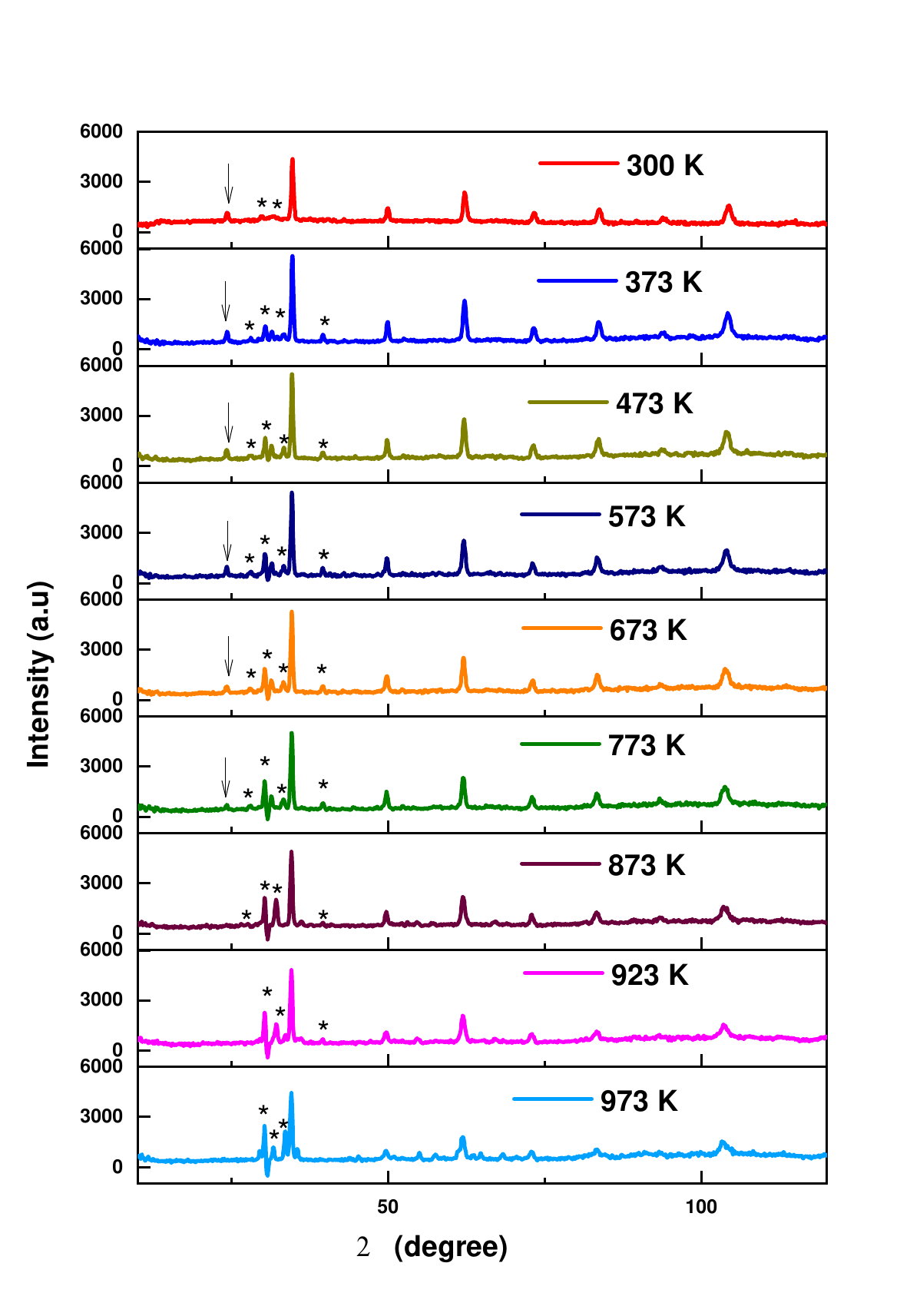}
	\caption{The NPD data of Cr$_3$Al pollycrystalline powder with furnace, collected with neutrons of $\lambda$= 1.2443 $\AA$  at T = 300, 373, 473, 573, 673, 773, 873, 923 and 973 K, from top to bottom, respectively. The black arrow represent the magnetic Bragg peak and the black asterisks tag indicates the peaks from the sample environment (furnace).}
	\label{fi2}
\end{figure}
The temperature-dependent NPD data reveal the evolution of the magnetic Bragg peak (200), which completely disappears between 773 K and 873 K. The impurity peaks, marked by black asterisks, originate from the sample environment (furnace). It can be clearly seen that the intensity of those peaks increases with increasing temperature, while the intensity of the magnetic Bragg peak gradually decreases and disappears above  $T_C$. 
\subsection{Magnetoresistance
}
\begin{figure}[H]
	\centering
	\hspace*{-0.6cm}\includegraphics[width=1.15\linewidth]{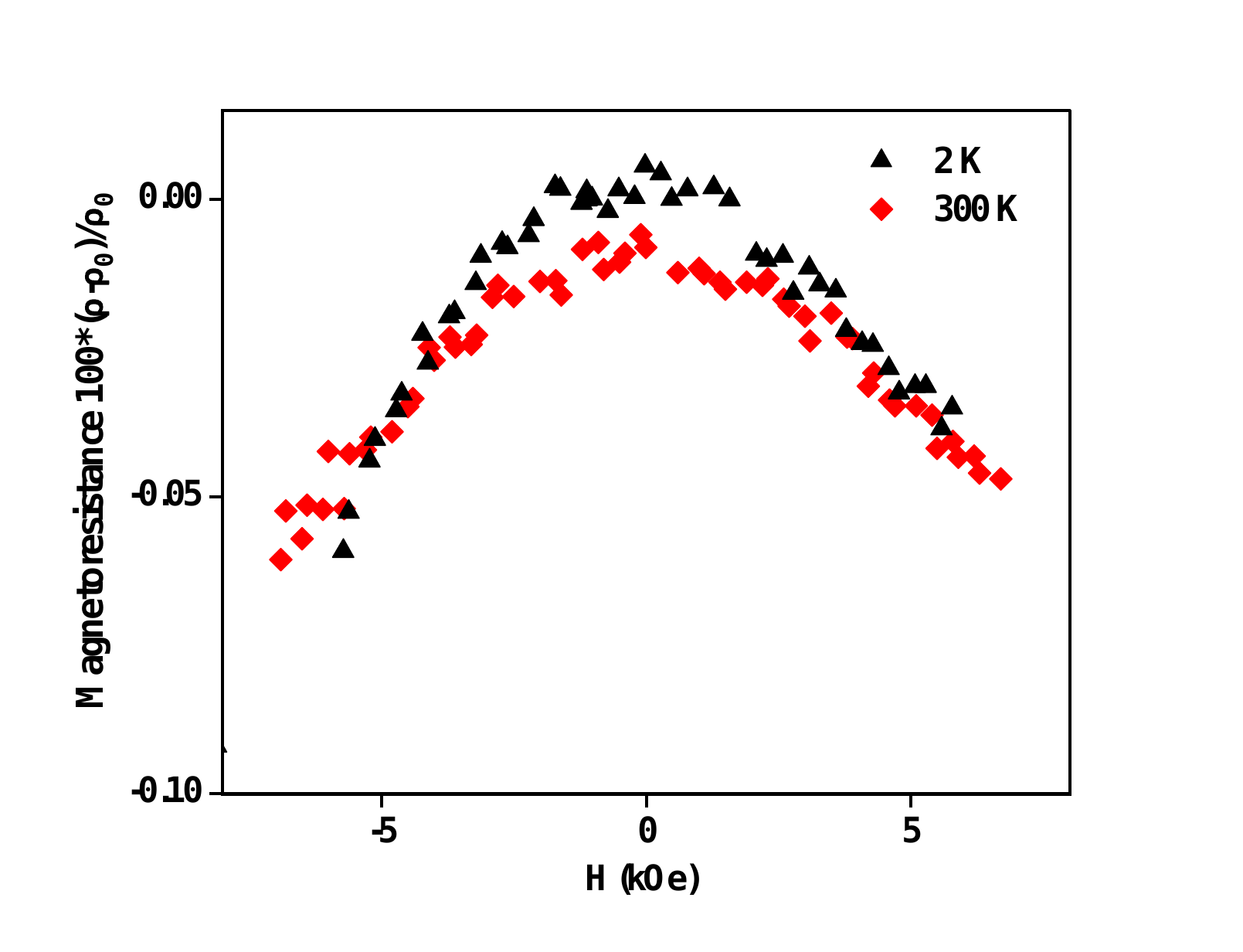}
	\caption{ Magnetoresistance of Cr$_3$Al measured at 2 K and 300 K.}
	\label{mr}
\end{figure}
The field-dependent magnetoresistance of Cr$_3$Al as given in Figure \ref{mr} exhibits a small negative MR that is nearly temperature-independent between 2 K and 300 K. The symmetric, non-hysteretic MR response is characteristic of spin-gapless semiconducting behavior, where field-induced spin alignment suppresses spin-disorder scattering without significant contribution from thermally activated carriers.
\subsection{Single-crystal X-ray diffraction (SCXRD) data}
\begin{table*}[t]
\centering
\caption{Crystallographic data for Cr$_3$Al single crystal}
\label{single crystal xrd}

\begin{tabular}{|l|l|}
\hline
\multicolumn{2}{|c|}{\textbf{Crystallographic data}} \\
\hline
EDS composition (at.\ \%) & Cr\textsubscript{73(3)}Al\textsubscript{27(3)} \\
$M_r$ & 90.5 \\
Crystal system, space group & Cubic, \textit{Im$\bar{3}$m} \\
Temperature (K) & 293 \\
$a$ (\AA) & 2.9422(8) \\
$V$ (\AA$^3$) & 25.47(1) \\
$Z$ & 1 \\
Radiation type & Mo K$\alpha$ \\
$\mu$ (mm\textsuperscript{$-$1}) & 15.28 \\
\hline
\multicolumn{2}{|c|}{\textbf{Data collection}} \\
\hline
Diffractometer & Bruker D8 Quest using Photon II detector \\
No. of measured, independent and observed [I $>$ 3$\sigma$(I)] reflections & 233, 13, 13 \\
$R_{\mathrm{int}}$ & 0.035 \\
$(\sin\theta / \lambda)_{\mathrm{max}}$ (\AA$^{-1}$) & 0.833 \\
\hline
\multicolumn{2}{|c|}{\textbf{Refinement}} \\
\hline
$R$[$F^2 > 2\sigma(F^2)$], $wR(F^2)$, $S$ & 0.008, 0.024, 1.12 \\
No. of reflections & 13 \\
No. of parameters & 3 \\
$\Delta \rho_{\mathrm{max}}, \Delta \rho_{\mathrm{min}}$ (e\,\AA$^{-3}$) & 0.15, -0.26 \\
\hline
\end{tabular}
\end{table*}

\begin{table}[H]
\centering
\small                                
\setlength{\tabcolsep}{3pt}           
\renewcommand{\arraystretch}{0.9}     

\caption{Atomic coordinates, refined atomic displacement parameters and site occupancy factor of Cr$_3$Al}

\begin{tabular}{|c|c|c|c|c|c|c|c|}
\hline
\textbf{Atom} & \textbf{Wyck.} & \textbf{Site} & \textbf{S.O.F.} & \textbf{x/a} & \textbf{y/b} & \textbf{z/c} & \textbf{U [\AA$^2$]} \\
\hline
Cr1/Al1 & 2\textit{a} & $m\bar{3}m$ & 0.73/0.27 & 0 & 0 & 0 & 0.0090(2) \\
\hline
\end{tabular}
\end{table}

\newpage

\bibliographystyle{unsrt}
\bibliography{references}

@article{Tavares2023,
title = {Heusler alloys: Past, properties, new alloys, and prospects},
journal = {Progress in Materials Science},
volume = {132},
pages = {101017},
year = {2023},
issn = {0079-6425},
doi = {https://doi.org/10.1016/j.pmatsci.2022.101017},
url = {https://www.sciencedirect.com/science/article/pii/S0079642522000986},
author = {Sheron Tavares and Kesong Yang and Marc A. Meyers}
}

@article{Groot1983,
  title = {New Class of Materials: Half-Metallic Ferromagnets},
  author = {de Groot, R. A. and Mueller, F. M. and Engen, P. G. van and Buschow, K. H. J.},
  journal = {Phys. Rev. Lett.},
  volume = {50},
  issue = {25},
  pages = {2024--2027},
  numpages = {0},
  year = {1983},
  month = {Jun},
  publisher = {American Physical Society},
  doi = {10.1103/PhysRevLett.50.2024},
  url = {https://link.aps.org/doi/10.1103/PhysRevLett.50.2024}
}

@article{Alijani2011,
  title = {Electronic, structural, and magnetic properties of the half-metallic ferromagnetic quaternary Heusler compounds {CoFeMnZ (Z= Al, Ga, Si, Ge) }},
  author = {Alijani, Vajiheh and Ouardi, Siham and Fecher, Gerhard H. and Winterlik, J\"urgen and Naghavi, S. Shahab and Kozina, Xeniya and Stryganyuk, Gregory and Felser, Claudia and Ikenaga, Eiji and Yamashita, Yoshiyuki and Ueda, Shigenori and Kobayashi, Keisuke},
  journal = {Phys. Rev. B},
  volume = {84},
  issue = {22},
  pages = {224416},
  numpages = {10},
  year = {2011},
  month = {Dec},
  publisher = {American Physical Society},
  doi = {10.1103/PhysRevB.84.224416},
  url = {https://link.aps.org/doi/10.1103/PhysRevB.84.224416}
}

@article{Seema2019,
title = {The effect of pressure and disorder on half-metallicity of {CoRuFeSi} Quaternary Heusler alloy},
journal = {Intermetallics},
volume = {110},
pages = {106478},
year = {2019},
issn = {0966-9795},
doi = {https://doi.org/10.1016/j.intermet.2019.106478},
url = {https://www.sciencedirect.com/science/article/pii/S0966979518302826},
author = {K. Seema},
}

@article{Wang2020,
title = {Spin-gapless semiconductors for future spintronics and electronics},
journal = {Physics Reports},
volume = {888},
pages = {1-57},
year = {2020},
issn = {0370-1573},
doi = {https://doi.org/10.1016/j.physrep.2020.08.004},
url = {https://www.sciencedirect.com/science/article/pii/S037015732030291X},
author = {Xiaotian Wang and Zhenxiang Cheng and Gang Zhang and Hongkuan Yuan and Hong Chen and Xiao-Lin Wang}
}

@article{Xia2022,
title = {Magnetic and anomalous transport properties in spin-gapless semiconductor like quaternary Heusler alloy {CoFeTiSn}},
journal = {Journal of Magnetism and Magnetic Materials},
volume = {553},
pages = {169283},
year = {2022},
issn = {0304-8853},
doi = {https://doi.org/10.1016/j.jmmm.2022.169283},
url = {https://www.sciencedirect.com/science/article/pii/S0304885322002311},
author = {Zhonghao Xia and Zhuhong Liu and Qiangqiang Zhang and Yajiu Zhang and Xingqiao Ma},
}

@Article{Gupta2024,
author ="Gupta, Shuvankar and Sau, Jyotirmoy and Kumar, Manoranjan and Mazumdar, Chandan",
title  ="Spin-gapless semiconducting characteristics and related band topology of quaternary Heusler alloy {CoFeMnSn}",
journal  ="J. Mater. Chem. C",
year  ="2024",
volume  ="12",
issue  ="2",
pages  ="706-716",
publisher  ="The Royal Society of Chemistry",
doi  ="10.1039/D3TC03481J",
url  ="http://dx.doi.org/10.1039/D3TC03481J",
}

@article{Venkateswara_2023,
  title = {{FeRhCrSi}: Spin semimetal with spin-valve behavior at room temperature},
  author = {Venkateswara, Y. and Nag, Jadupati and Samatham, S. Shanmukharao and Patel, Akhilesh Kumar and Babu, P. D. and Varma, Manoj Raama and Nayak, Jayita and Suresh, K. G. and Alam, Aftab},
  journal = {Phys. Rev. B},
  volume = {107},
  issue = {10},
  pages = {L100401},
  numpages = {5},
  year = {2023},
  month = {Mar},
  publisher = {American Physical Society},
  doi = {10.1103/PhysRevB.107.L100401},
  url = {https://link.aps.org/doi/10.1103/PhysRevB.107.L100401}
}

@article{Harikrishnan2023,
  title = {Spin semimetallic behavior and sublattice spin crossover in the fully compensated ferrimagnetic half-Heusler compound {$({\mathrm{Co}}_{0.5}{\mathrm{Mn}}_{0.5})\mathrm{MnAl}$}},
  author = {Harikrishnan, R. and Bidika, Jatin Kumar and Nanda, B. R. K. and Chelvane, Arout J. and Kaushik, S. D. and Babu, P. D. and Narayanan, Harish Kumar},
  journal = {Phys. Rev. B},
  volume = {108},
  issue = {9},
  pages = {094407},
  numpages = {12},
  year = {2023},
  month = {Sep},
  publisher = {American Physical Society},
  doi = {10.1103/PhysRevB.108.094407},
  url = {https://link.aps.org/doi/10.1103/PhysRevB.108.094407}
}

@article{Xu_2015,
doi = {10.1209/0295-5075/111/68003},
url = {https://dx.doi.org/10.1209/0295-5075/111/68003},
year = {2015},
month = {oct},
publisher = {EDP Sciences, IOP Publishing and Società Italiana di Fisica},
volume = {111},
number = {6},
pages = {68003},
author = {Xu, G. Z. and Zhang, X. M. and Hou, Z. P. and Wang, Y. and Liu, E. K. and Xi, X. K. and Wang, S. G. and Wang, W. Q. and Luo, H. Z. and Wang, W. H. and Wu, G. H.},
title = {New spin injection scheme based on spin gapless semiconductors: A first-principles study},
journal = {Europhysics Letters},
}

@Article{Wederni2024,
AUTHOR = {Wederni, Asma and Daza, Jason and Ben Mbarek, Wael and Saurina, Joan and Escoda, Lluisa and Sunol, Joan-Josep},
TITLE = {Crystal Structure and Properties of Heusler Alloys: A Comprehensive Review},
JOURNAL = {Metals},
VOLUME = {14},
YEAR = {2024},
NUMBER = {6},
ARTICLE-NUMBER = {688},
URL = {https://www.mdpi.com/2075-4701/14/6/688},
ISSN = {2075-4701},
DOI = {10.3390/met14060688}
}

@article{Philip2024,
author = {Philip, Reshna Elsa and Kumar, Rahul and Manni, S.},
title = {Structural, magnetic and electrical properties of Heusler alloy {Cr$_3$Al}},
journal = {AIP Conference Proceedings},
volume = {2995},
number = {1},
pages = {020151},
year = {2024},
month = {01},
issn = {0094-243X},
doi = {10.1063/5.0178030},
url = {https://doi.org/10.1063/5.0178030},
}

@article{Wang2016,
title = {Origin of d$^0$ half-metallic characteristic in {DO$_3$-type XO$_3$} {(X=Li, Na, K and Rb) compounds}},
journal = {Journal of Magnetism and Magnetic Materials},
volume = {412},
pages = {95-101},
year = {2016},
issn = {0304-8853},
doi = {https://doi.org/10.1016/j.jmmm.2016.04.006},
url = {https://www.sciencedirect.com/science/article/pii/S0304885316303080},
author = {Xiaotian Wang and Zhenxiang Cheng and Jianli Wang and Habib Rozale and Juntao Yang and Zheyin Yu and Guodong Liu},
}

@article{Zhang2012,
title = {Electronic structures and magnetism of {Cr$_3$Z (Z=Si, Ge, Sb) with DO$_3$ structures}},
journal = {Computational Materials Science},
volume = {65},
pages = {456-460},
year = {2012},
issn = {0927-0256},
doi = {https://doi.org/10.1016/j.commatsci.2012.08.018},
url = {https://www.sciencedirect.com/science/article/pii/S0927025612005101},
author = {X.M. Zhang and X.F. Dai and H.Y. Jia and G.F. Chen and H.Y. Liu and H.Z. Luo and Y. Li and X. Yu and G.D. Liu and W.H. Wang and G.H. Wu},
}

@article{Zhang2013,
doi = {10.1088/0953-8984/25/20/206006},
url = {https://dx.doi.org/10.1088/0953-8984/25/20/206006},
year = {2013},
month = {apr},
publisher = {IOP Publishing},
volume = {25},
number = {20},
pages = {206006},
author = {Zhang, Delin and Yan, Binghai and Wu, Shu-Chun and Kübler, Jürgen and Kreiner, Guido and Parkin, Stuart S P and Felser, Claudia},
title = {First-principles study of the structural stability of cubic, tetragonal and hexagonal phases in {Mn$_3Z$ (Z=Ga, Sn and Ge)} Heusler compounds},
journal = {Journal of Physics: Condensed Matter},
}

@article{Jeong2012,
title = {Magnetic properties of {Mn$_3$Si} from first-principles studies},
journal = {Physica B: Condensed Matter},
volume = {407},
number = {5},
pages = {888-890},
year = {2012},
issn = {0921-4526},
doi = {https://doi.org/10.1016/j.physb.2011.12.110},
url = {https://www.sciencedirect.com/science/article/pii/S0921452611013032},
author = {T. Jeong},
}

@article{Erkisi2018, 
title={The Investigation {DO$_3$-type Fe$_3$M (M=Al, Ga, Si and Ge)} Full-Heusler Alloys Within First Principles Study},
journal={Politeknik Dergisi},
volume={21}, 
pages={927–936}, 
year={2018}, 
DOI={10.2339/politeknik.385443}, 
author={Erkisi, Aytac and Surucu, Gokhan}, 
}

@article{Chattaraj2022,
title = {Ab initio study of the structural stability and Dirac fermion behaviour in {A$_3$B (A = Cr, Mo, W; B = Al, Ga, In, Si, Ge, Sn, Be) and W$_3$M, (M = Ru, Ta, Re, Os, Ir, Au) compounds}},
journal = {Physica B: Condensed Matter},
volume = {646},
pages = {414315},
year = {2022},
issn = {0921-4526},
doi = {https://doi.org/10.1016/j.physb.2022.414315},
url = {https://www.sciencedirect.com/science/article/pii/S0921452622006007},
author = {Ananya Chattaraj and Aloke Kanjilal and Vijay Kumar},
}

@article{Li2009,
author = {Li, Jia and Chen, Hongjian and Li, Yangxian and Xiao, Yu and Li, Zhiqing},
title = {A theoretical design of half-metallic compounds by a long range of doping {Mn for Heusler-type Cr$_3$Al}},
journal = {Journal of Applied Physics},
volume = {105},
number = {8},
pages = {083717},
year = {2009},
month = {04},
issn = {0021-8979},
doi = {10.1063/1.3116533},
url = {https://doi.org/10.1063/1.3116533},
}

@article{Boekelheide2012Aug,
title = {Chemical ordering in {Cr${}_{3}$Al} and relation to semiconducting behavior},
author = {Boekelheide, Z. and Stewart, D. A. and Hellman, F.},
journal = {Phys. Rev. B},
volume = {86},
issue = {8},
pages = {085120},
numpages = {10},
year = {2012},
month = {Aug},
publisher = {American Physical Society},
doi = {10.1103/PhysRevB.86.085120},
url = {https://link.aps.org/doi/10.1103/PhysRevB.86.085120}
}

@article{Boekelheide2012Mar,
 title = {Antiferromagnetism in {Cr${}_{3}$Al} and relation to semiconducting behavior},
author = {Boekelheide, Z. and Saerbeck, T. and Stampfl, Anton P. J. and Robinson, R. A. and Stewart, D. A. and Hellman, F.},
journal = {Phys. Rev. B},
volume = {85},
issue = {9},
pages = {094413},
numpages = {8},
year = {2012},
month = {Mar},
publisher = {American Physical Society},
doi = {10.1103/PhysRevB.85.094413},
url = {https://link.aps.org/doi/10.1103/PhysRevB.85.094413}
}

@article{TOYOKI2021,
title = {Dominant carrier of pseudo-gap antiferromagnet {Cr$_3$Al} thin film},
journal = {Physica B: Condensed Matter},
volume = {620},
pages = {413281},
year = {2021},
issn = {0921-4526},
doi = {https://doi.org/10.1016/j.physb.2021.413281},
url = {https://www.sciencedirect.com/science/article/pii/S0921452621004543},
author = {Kentaro Toyoki and Masayuki Hayashi and Shunsuke Hamaguchi and Noriaki Kishida and Yu Shiratsuchi and Takafumi Ishibe and Yoshiaki Nakamura and Ryoichi Nakatani},
}

@article{Chakrabarti1971,
title={Transport properties of {Cr-Al} solid solutions},
author={Debalay Chakrabarti, P.A. Beck},
journal={Elsevier ScienceDirect Journals},
year={1971},
volume={32},
number={7},
doi={10.1016/s0022-3697(71)80054-9}
}

@article{DENBroeder1981,
title = "Microstructure of {Cr$_{100-x}$Al$_x$} alloys studied by means of transmission electron microscopy and diffraction. II. Discovery of a new phase",
author = "{den Broeder}, {F. J.A.} and {van Tendeloo}, G. and S. Amelinckx and J. Hornstra and {de Ridder}, R. and {van Landuyt}, J. and {van Daal}, {H. J.}",
year = "1981",
month = "sep",
day = "16",
doi = "10.1002/pssa.2210670123",
language = "English",
volume = "67",
pages = "233-248",
journal = "physica status solidi (a)",
issn = "0031-8965",
publisher = "Wiley - John Wiley & Sons, Ltd",
number = "1",
}

@article{Bainsla2017,
title = {Structural and magnetic properties of epitaxial thin films of the equiatomic quaternary {CoFeMnSi} Heusler alloy},
author = {Bainsla, Lakhan and Yilgin, Resul and Okabayashi, Jun and Ono, Atsuo and Suzuki, Kazuya and Mizukami, Shigemi},
journal = {Phys. Rev. B},
volume = {96},
issue = {9},
pages = {094404},
numpages = {9},
year = {2017},
month = {Sep},
publisher = {American Physical Society},
doi = {10.1103/PhysRevB.96.094404},
url = {https://link.aps.org/doi/10.1103/PhysRevB.96.094404}
}

@article{Slater1936,
title = {The Ferromagnetism of Nickel},
author = {Slater, J. C.},
journal = {Phys. Rev.},
volume = {49},
issue = {7},
pages = {537--545},
numpages = {0},
year = {1936},
month = {Apr},
publisher = {American Physical Society},
doi = {10.1103/PhysRev.49.537},
url = {https://link.aps.org/doi/10.1103/PhysRev.49.537}
}

@article{Pauling1938,
title = {The Nature of the Interatomic Forces in Metals},
author = {Pauling, Linus},
journal = {Phys. Rev.},
volume = {54},
issue = {11},
pages = {899--904},
numpages = {0},
year = {1938},
month = {Dec},
publisher = {American Physical Society},
doi = {10.1103/PhysRev.54.899},
url = {https://link.aps.org/doi/10.1103/PhysRev.54.899}
}

@article{Venkateswara2018,
title = {Competing magnetic and spin-gapless semiconducting behavior in fully compensated ferrimagnetic {CrVTiAl}: Theory and experiment},
author = {Venkateswara, Y. and Gupta, Sachin and Samatham, S. Shanmukharao and Varma, Manoj Raama and Enamullah and Suresh, K. G. and Alam, Aftab},
journal = {Phys. Rev. B},
volume = {97},
issue = {5},
pages = {054407},
numpages = {8},
year = {2018},
month = {Feb},
publisher = {American Physical Society},
doi = {10.1103/PhysRevB.97.054407},
url = {https://link.aps.org/doi/10.1103/PhysRevB.97.054407}
}

@article{Rani2019,
title = {Spin-gapless semiconducting nature of {Co}-rich {${\mathrm{Co}}_{1+\mathit{x}}{\mathrm{Fe}}_{1\ensuremath{-}\mathit{x}}\mathrm{CrGa}$}},
author = {Rani, Deepika and Enamullah and Bainsla, Lakhan and Suresh, K. G. and Alam, Aftab},
journal = {Phys. Rev. B},
volume = {99},
issue = {10},
pages = {104429},
numpages = {10},
year = {2019},
month = {Mar},
publisher = {American Physical Society},
doi = {10.1103/PhysRevB.99.104429},
url = {https://link.aps.org/doi/10.1103/PhysRevB.99.104429}
}

@article{Bainsla2015,
title = {Spin gapless semiconducting behavior in equiatomic quaternary {CoFeMnSi} Heusler alloy},
author = {Bainsla, Lakhan and Mallick, A. I. and Raja, M. Manivel and Nigam, A. K. and Varaprasad, B. S. D. Ch. S. and Takahashi, Y. K. and Alam, Aftab and Suresh, K. G. and Hono, K.},
journal = {Phys. Rev. B},
volume = {91},
issue = {10},
pages = {104408},
numpages = {6},
year = {2015},
month = {Mar},
publisher = {American Physical Society},
doi = {10.1103/PhysRevB.91.104408},
url = {https://link.aps.org/doi/10.1103/PhysRevB.91.104408}
}

@article{Gao2013,
    author = {Gao, G. Y. and Yao, Kai-Lun},
    title = "{Antiferromagnetic half-metals, gapless half-metals, and spin gapless semiconductors: The DO3-type Heusler alloys}",
    journal = {Applied Physics Letters},
    volume = {103},
    number = {23},
    pages = {232409},
    year = {2013},
    month = {12},
    issn = {0003-6951},
    doi = {10.1063/1.4840318},
    url = {https://doi.org/10.1063/1.4840318},
}

@article{Zhao2019,
author = "Zhao, W. Q. and Dai, X. F. and Zhang, X. M. and Mo, Z. J. and Wang, X. T. and Chen, G. F. and Cheng, Z. X. and Liu, G. D.",
title = "{Preparation and physical properties of a Cr${\sb 3}$Al film with a DO${\sb 3}$ structure}",
journal = "IUCrJ",
year = "2019",
volume = "6",
number = "4",
pages = "552--557",
month = "Jul",
doi = {10.1107/S2052252519004469},
url = {https://doi.org/10.1107/S2052252519004469},
}

@article{34_Kresse1996,
title = {Efficient iterative schemes for ab initio total-energy calculations using a plane-wave basis set},
author = {Kresse, G. and Furthm\"uller, J.},
journal = {Phys. Rev. B},
volume = {54},
issue = {16},
pages = {11169--11186},
numpages = {0},
year = {1996},
month = {Oct},
publisher = {American Physical Society},
doi = {10.1103/PhysRevB.54.11169},
url = {https://link.aps.org/doi/10.1103/PhysRevB.54.11169}
}

@article{35_Kresse1996,
title = {Efficiency of ab-initio total energy calculations for metals and semiconductors using a plane-wave basis set},
journal = {Computational Materials Science},
volume = {6},
number = {1},
pages = {15-50},
year = {1996},
issn = {0927-0256},
doi = {https://doi.org/10.1016/0927-0256(96)00008-0},
url = {https://www.sciencedirect.com/science/article/pii/0927025696000080},
author = {G. Kresse and J. Furthmüller},
}

@article{36_Kresse1993,
title = {Ab initio molecular dynamics for liquid metals},
author = {Kresse, G. and Hafner, J.},
journal = {Phys. Rev. B},
volume = {47},
issue = {1},
pages = {558--561},
numpages = {0},
year = {1993},
month = {Jan},
publisher = {American Physical Society},
doi = {10.1103/PhysRevB.47.558},
url = {https://link.aps.org/doi/10.1103/PhysRevB.47.558}
}

@article{37_Kresse1999,
title = {From ultrasoft pseudopotentials to the projector augmented-wave method},
author = {Kresse, G. and Joubert, D.},
journal = {Phys. Rev. B},
volume = {59},
issue = {3},
pages = {1758--1775},
numpages = {0},
year = {1999},
month = {Jan},
publisher = {American Physical Society},
doi = {10.1103/PhysRevB.59.1758},
url = {https://link.aps.org/doi/10.1103/PhysRevB.59.1758}
}

@article{38_Sun2015,
title = {Strongly Constrained and Appropriately Normed Semilocal Density Functional},
author = {Sun, Jianwei and Ruzsinszky, Adrienn and Perdew, John P.},
journal = {Phys. Rev. Lett.},
volume = {115},
issue = {3},
pages = {036402},
numpages = {6},
year = {2015},
month = {Jul},
publisher = {American Physical Society},
doi = {10.1103/PhysRevLett.115.036402},
url = {https://link.aps.org/doi/10.1103/PhysRevLett.115.036402}
}

@article{39_Monkhorst1976,
title = {Special points for Brillouin-zone integrations},
author = {Monkhorst, Hendrik J. and Pack, James D.},
journal = {Phys. Rev. B},
volume = {13},
issue = {12},
pages = {5188--5192},
numpages = {0},
year = {1976},
month = {Jun},
publisher = {American Physical Society},
doi = {10.1103/PhysRevB.13.5188},
url = {https://link.aps.org/doi/10.1103/PhysRevB.13.5188}
}

@article{40_Blochl1994,
title = {Improved tetrahedron method for Brillouin-zone integrations},
author = {Bl\"ochl, Peter E. and Jepsen, O. and Andersen, O. K.},
journal = {Phys. Rev. B},
volume = {49},
issue = {23},
pages = {16223--16233},
numpages = {0},
year = {1994},
month = {Jun},
publisher = {American Physical Society},
doi = {10.1103/PhysRevB.49.16223},
url = {https://link.aps.org/doi/10.1103/PhysRevB.49.16223}
}

@article{41_Zunger1990,
title = {Special quasirandom structures},
author = {Zunger, Alex and Wei, S.-H. and Ferreira, L. G. and Bernard, James E.},
journal = {Phys. Rev. Lett.},
volume = {65},
issue = {3},
pages = {353--356},
numpages = {0},
year = {1990},
month = {Jul},
publisher = {American Physical Society},
doi = {10.1103/PhysRevLett.65.353},
url = {https://link.aps.org/doi/10.1103/PhysRevLett.65.353}
}

@article{42_Vandewalle2013,
title = {Efficient stochastic generation of special quasirandom structures},
journal = {Calphad},
volume = {42},
pages = {13-18},
year = {2013},
issn = {0364-5916},
doi = {https://doi.org/10.1016/j.calphad.2013.06.006},
url = {https://www.sciencedirect.com/science/article/pii/S0364591613000540},
author = {A. {van de Walle} and P. Tiwary and M. {de Jong} and D.L. Olmsted and M. Asta and A. Dick and D. Shin and Y. Wang and L.-Q. Chen and Z.-K. Liu},
}

@article{44_Ouardi2013,
title = {Realization of Spin Gapless Semiconductors: The Heusler Compound {${\mathrm{Mn}}_{2}\mathrm{CoAl}$}},
author = {Ouardi, Siham and Fecher, Gerhard H. and Felser, Claudia and K\"ubler, J\"urgen},
journal = {Phys. Rev. Lett.},
volume = {110},
issue = {10},
pages = {100401},
numpages = {5},
year = {2013},
month = {Mar},
publisher = {American Physical Society},
doi = {10.1103/PhysRevLett.110.100401},
url = {https://link.aps.org/doi/10.1103/PhysRevLett.110.100401}
}

@article{Webster1969,
author = {Peter J. Webster},
title = {Heusler alloys},
journal = {Contemporary Physics},
volume = {10},
number = {6},
pages = {559--577},
year = {1969},
publisher = {Taylor \& Francis},
doi = {10.1080/00107516908204800},
URL = {https://doi.org/10.1080/00107516908204800
},

}

@article{Jamer2017,
  title = {Compensated Ferrimagnetism in the Zero-Moment Heusler Alloy {${\mathrm{Mn}}_{3}\mathrm{Al}$}},
  author = {Jamer, Michelle E. and Wang, Yung Jui and Stephen, Gregory M. and McDonald, Ian J. and Grutter, Alexander J. and Sterbinsky, George E. and Arena, Dario A. and Borchers, Julie A. and Kirby, Brian J. and Lewis, Laura H. and Barbiellini, Bernardo and Bansil, Arun and Heiman, Don},
  journal = {Phys. Rev. Appl.},
  volume = {7},
  issue = {6},
  pages = {064036},
  numpages = {7},
  year = {2017},
  month = {Jun},
  publisher = {American Physical Society},
  doi = {10.1103/PhysRevApplied.7.064036},
  url = {https://link.aps.org/doi/10.1103/PhysRevApplied.7.064036}
}

@book{book,
author = {Kittel2007},
year = {2007},
month = {01},
pages = {688},
title = {Introduction to Solid State Physics, 7th Ed},
isbn = {9788126510450}
}

@misc{Reitveld,
  howpublished ={{https://www.ill.eu/sites/fullprof/php/tutorials.html}},
}

@ARTICLE{Lind1984,
	author = {Lind, M.A.},
	title = {Measurements of the Temperature Dependence of Energy Gaps in Magnetically Ordered {Cr-Rich Cr-Al} Alloys},
	year = {1984},
	journal = {Journal of the Physical Society of Japan},
	volume = {53},
	number = {11},
	pages = {4029 – 4043},
	doi = {10.1143/JPSJ.53.4029},
	url = {https://www.scopus.com/inward/record.uri?eid=2-s2.0-0021521229&doi=10.1143%2fJPSJ.53.4029&partnerID=40&md5=77793578c1e863446936dee2a6f8151d},
	type = {Article},
	publication_stage = {Final},
	source = {Scopus}
}

@article{gupt2023,
    author = {Dutt Gupt, Guru and Kareri, Yousef and Hester, James and Ulrich, Clemens and Dhaka, R. S.},
    title = {Unveiling the magnetic structure and phase transition of {${\mathrm{Cr}}_{2}\mathrm{CoAl}$} using neutron diffraction},
    journal = {Journal of Applied Physics},
    volume = {134},
    number = {11},
    pages = {113901},
    year = {2023},
    month = {09},
    issn = {0021-8979},
    doi = {10.1063/5.0158138},
    url = {https://doi.org/10.1063/5.0158138},  
}

@article{shukla2019,
    author = {Ahad, Abdul and Shukla, D. K.},
    title = {A setup for Seebeck coefficient measurement through controlled heat pulses},
    journal = {Review of Scientific Instruments},
    volume = {90},
    number = {11},
    pages = {116101},
    year = {2019},
    month = {11},
    issn = {0034-6748},
    doi = {10.1063/1.5116160},
    url = {https://doi.org/10.1063/1.5116160}
}

@article{superflip,
author = "Palatinus, Luk{\'{a}}{\v{s}} and Chapuis, Gervais",
title = "{{\it SUPERFLIP} {--} a computer program for the solution of crystal structures by charge flipping in arbitrary dimensions}",
journal = "Journal of Applied Crystallography",
year = "2007",
volume = "40",
number = "4",
pages = "786--790",
month = "Aug",
doi = {10.1107/S0021889807029238},
url = {https://doi.org/10.1107/S0021889807029238},
}

@article{jana2006,
author = {Petrícek, Václav and Dusek, Michal},
year = {2006},
month = {01},
pages = {},
title = {The crystallographic computing system},
journal = {Institute of physics, Praha, Czech Republic}
}

@article{Rani2020,
    author = {Rani, Deepika and Bainsla, Lakhan and Alam, Aftab and Suresh, K. G.},
    title = {Spin-gapless semiconductors: Fundamental and applied aspects},
    journal = {Journal of Applied Physics},
    volume = {128},
    number = {22},
    pages = {220902},
    year = {2020},
    month = {12},
    issn = {0021-8979},
    doi = {10.1063/5.0028918},
    url = {https://doi.org/10.1063/5.0028918},
}

@misc{APEX5,
  title        = {APEX5 software},
  author       = {{Bruker AXS Inc.}},
  address      = {Madison, WI},
  year         = {2023},
  note         = {Bruker AXS Inc., Madison, WI (2023)}
}

@article{Bainsla2015CoFecRGa,
  title = {Origin of spin gapless semiconductor behavior in {CoFeCrGa}: Theory and Experiment},
  author = {Bainsla, Lakhan and Mallick, A. I. and Raja, M. Manivel and Coelho, A. A. and Nigam, A. K. and Johnson, D. D. and Alam, Aftab and Suresh, K. G.},
  journal = {Phys. Rev. B},
  volume = {92},
  issue = {4},
  pages = {045201},
  numpages = {5},
  year = {2015},
  month = {Jul},
  publisher = {American Physical Society},
  doi = {10.1103/PhysRevB.92.045201},
  url = {https://link.aps.org/doi/10.1103/PhysRevB.92.045201}
}

@article{wang2008,
  title = {Proposal for a New Class of Materials: Spin Gapless Semiconductors},
  author = {Wang, X. L.},
  journal = {Phys. Rev. Lett.},
  volume = {100},
  issue = {15},
  pages = {156404},
  numpages = {4},
  year = {2008},
  month = {Apr},
  publisher = {American Physical Society},
  doi = {10.1103/PhysRevLett.100.156404},
  url = {https://link.aps.org/doi/10.1103/PhysRevLett.100.156404}
}

@article{Zhu2024, doi = {10.21105/joss.05974}, url = {https://doi.org/10.21105/joss.05974}, year = {2024}, publisher = {The Open Journal}, volume = {9}, number = {93}, pages = {5974}, author = {Zhu, Bonan and Kavanagh, Seán R. and Scanlon, David}, title = {easyunfold: A Python package for unfolding electronic band structures}, journal = {Journal of Open Source Software} }

@article{Mishra2023,
  title        = {Disordered spin gapless semiconducting {CoFeCrGa} Heusler alloy thin films on {Si}(100): experiment and theory},
  author       = {Vireshwar Mishra and Amar Kumar and Lalit Pandey and Nanhe Kumar Gupta and Soumyarup Hait and Vineet Barwal and Nikita Sharma and Nakul Kumar and Sharat Chandra and Sujeet Chaudhary},
  journal      = {Nanoscale},
  year         = {2023},
  volume       = {15},
  number       = {1},
  pages        = {337--349},
  doi          = {10.1039/D2NR03424G},
  url          = {https://pubs.rsc.org/en/content/articlehtml/2023/nr/d2nr03424g}
}

@article{huarui2019,
    title={Structures, magnetism and transport properties of the potential spin-gapless semiconductor CoFeMnSi alloy},
    author={Fu, Huarui and Li, Yunlong and Ma, Li and You, Caiyin and Zhang, Qing and Tian, Na},
    volume={473},
    url={http://dx.doi.org/10.1016/j.jmmm.2018.10.040},
    doi={10.1016/j.jmmm.2018.10.040},
    journal={Journal of Magnetism and Magnetic Materials},
    publisher={Elsevier BV},
    year={2019},
    month={Mar},
    pages={16-20},
    language={en},
}

@article{Xu2014,
    author = {Xu, G. Z. and Du, Y. and Zhang, X. M. and Zhang, H. G. and Liu, E. K. and Wang, W. H. and Wu, G. H.},
    title = {Magneto-transport properties of oriented Mn2CoAl films sputtered on thermally oxidized Si substrates},
    journal = {Applied Physics Letters},
    volume = {104},
    number = {24},
    pages = {242408},
    year = {2014},
    month = {06},
    issn = {0003-6951},
    doi = {10.1063/1.4884203},
    url = {https://doi.org/10.1063/1.4884203},
    eprint = {https://pubs.aip.org/aip/apl/article-pdf/doi/10.1063/1.4884203/13152187/242408_1_online.pdf},
}

@article{VANDEWALLE2013,
title = {Efficient stochastic generation of special quasirandom structures},
journal = {Calphad},
volume = {42},
pages = {13-18},
year = {2013},
issn = {0364-5916},
doi = {https://doi.org/10.1016/j.calphad.2013.06.006},
url = {https://www.sciencedirect.com/science/article/pii/S0364591613000540},
author = {A. {van de Walle} and P. Tiwary and M. {de Jong} and D.L. Olmsted and M. Asta and A. Dick and D. Shin and Y. Wang and L.-Q. Chen and Z.-K. Liu}
}

\end{document}